\documentclass[floats,floatfix,showpacs,amssymb,prd,twocolumn,superscriptaddress,reprint,nofootinbib,nolongbibliography]{revtex4-2}

\usepackage{amssymb,amsmath,verbatim,mathtools,needspace,enumitem,etoolbox,graphicx,physics,microtype,afterpage,bigints,gensymb,tabularx,xspace,bm}%

\usepackage{xfrac}
\usepackage[dvipsnames, usenames]{xcolor}
\definecolor{linkcolor}{rgb}{0.0,0.3,0.5}
\definecolor{dodgerblue}{HTML}{1E90FF}
\usepackage[unicode, colorlinks=true, linkcolor=linkcolor, citecolor=linkcolor, filecolor=linkcolor,urlcolor=linkcolor, pdfusetitle]{hyperref}
\usepackage[all]{hypcap}
\usepackage[T1]{fontenc}
\usepackage[utf8]{inputenc}
\usepackage{orcidlink}
\usepackage{tensor}
\makeatletter
\newsavebox{\@brx}
\newcommand{\llangle}[1][]{\savebox{\@brx}{\(\m@th{#1\langle}\)}%
  \mathopen{\copy\@brx\kern-0.6\wd\@brx\usebox{\@brx}}}
\newcommand{\rrangle}[1][]{\savebox{\@brx}{\(\m@th{#1\rangle}\)}%
  \mathclose{\copy\@brx\kern-0.6\wd\@brx\usebox{\@brx}}}
\makeatother

\renewcommand{\vec}[1]{\boldsymbol{#1}}
\usepackage{graphicx} %

\makeatletter
\newcommand*{\balancecolsandclearpage}{\close@column@grid \cleardoublepage \twocolumngrid}
\makeatother

\newcommand{\milan}{\affiliation{Dipartimento di Fisica ``G. Occhialini'', Universit\'a degli Studi di Milano-Bicocca, Piazza della Scienza 3, 20126 Milano, Italy}}
\newcommand{\infn}{\affiliation{INFN, Sezione di Milano-Bicocca, Piazza della Scienza 3, 20126 Milano, Italy}}

\begin{document}

\title{Non-adiabatic dynamics of eccentric black-hole \\  binaries in post-Newtonian theory}%

\author{Giulia Fumagalli$\,$\orcidlink{0009-0004-2044-989X}}
\email{g.fumagalli47@campus.unimib.it}
\milan \infn 
\author{Nicholas Loutrel$\,$\orcidlink{0000-0002-1597-3281}}
\email{nicholas.loutrel@unimib.it}
\milan \infn 
\author{Davide Gerosa$\,$\orcidlink{0000-0002-0933-3579}}
\milan \infn
\author{Matteo Boschini$\,$\orcidlink{0009-0002-5682-1871}}
\milan \infn


\date{\today}

\begin{abstract}

Eccentric black-hole binaries are among the most awaited sources of gravitational waves, yet their dynamics lack a consistent framework that provides a detailed and physically robust evolutionary description due to gauge issues. We present a new set of non-orbit-averaged equations, free from radiation-reaction gauge parameters, that accurately describe the evolution of orbital elements for eccentric, non-spinning black-hole binaries. We derive these equations by mapping the Keplerian orbital elements to a new set of characteristic parameters using energy and angular momentum definitions combined with near-identity transformations. The resulting framework is valid for arbitrary eccentricities, including parabolic and hyperbolic limits. %
Using this framework, we demonstrate %
 effects of the non-adiabatic emission of gravitational waves ---characteristic of eccentric binaries--- on the orbital parameters. Furthermore, we assess the regime of validity of the widely used orbit-averaged equations first derived by Peters in 1964. %
Importantly, their breakdown becomes evident at the first pericenter passage, implying that the validity of the orbit-averaged approximation cannot be inferred solely from binary initial conditions. %
The formalism we introduce, accurate up to 2.5 post-Newtonian order, aims to provide a robust tool for making reliable astrophysical predictions and accurately interpreting current and future gravitational wave data, paving the way for deeper insights into the dynamics of eccentric black hole binaries.
\end{abstract}

\maketitle

\section{Introduction}

\subsection{Context}

Accurately describing the dynamics of  non-spinning black holes (BHs) on eccentric orbits is a long-standing and fundamental challenge in astrophysics 
(see e.g.~\cite{DePorzio:2024cet, Zeeshan:2024ovp, DOrazio:2018jnv, Samsing:2018isx, Samsing:2018nxk}). 
This problem has been approached from various perspectives over more than sixty years due to its critical importance, particularly in the gravitational wave (GW) field~\cite{2019CQGra..36b5004L, 1985AIHPA..43..107D, LincolnWill, Damour:2004bz, Pound:2005fs, Shaikh:2023ypz, Boschini:2024scu}.
For instance, a wide variety of formation channels for binary systems, ranging from stellar-mass~\cite{2018PhRvD..97j3014S, 2021ApJ...907L..20T, Gondan:2020svr, Rom:2023kqm, DallAmico:2023neb, Osburn:2015duj, Romero-Shaw:2024klf, DePorzio:2024cet} to supermassive BHs~\cite{2013CQGra..30x4009S, 2022MNRAS.511.4753G}, predict their evolution on eccentric orbits.
This highlights the need to understand eccentric system dynamics and the long-term evolution of their orbital parameters, which is the first step toward building accurate waveform templates for GW detections.  
The dynamics of binary BHs is typically described using balance laws: the energy and angular momentum carried away by GWs are equated to the corresponding losses in the energy and angular momentum of the orbit, which directly leads to the coalescence of the binary~\cite{landau,gravitation, poisson,maggiore, Blanchet_lrr}.
The seminal work by Peters~\cite{1964PhRv..136.1224P} was among the first to apply balance laws to eccentric binaries, establishing the most widely used framework for their dynamics. This formalism employs a Newtonian description of the orbital energy and angular momentum, and leverages an orbit-average approximation to capture the secular evolution of the orbital elements~\cite{LincolnWill, Iyer:1993xi, Mora:2002gf}.

However, this approach has important limitations. The concept of averaging pre-dates Peters' work, and was originally proposed by Isaacson~\cite{1968PhRv..166.1263I, 1968PhRv..166.1272I} as a method for handling gauge effects in the definitions of energy and angular momentum from a suitably defined stress-energy tensor. Yet, this averaging procedure relies on the existence of a GW timescale that is longer than the timescale of gauge-dependent oscillations, allowing the deployment of multiple scale analysis~\cite{bender78:AMM} from which the secular (or averaged) approximation would constitute the leading order behavior. For generic binary systems, the latter is governed by the orbital timescale, while the former is the dissipative timescale associated with orbital energy loss due to GW emission, the so-called backreaction or radiation-reaction timescale.
In nature, the requirement of separation of scales is not always met.
Post-adiabatic corrections to the secular limit become important when constructing waveform templates \cite{2019CQGra..36aLT01L, 2019CQGra..36b5004L, Pound:2005fs, LincolnWill}. Further, for highly eccentric binaries,  GW emission is concentrated during pericenter passage~\cite{1977ApJ...216..610T}, and the timescale associated with the emission is thus shorter than the orbital period~\cite{Loutrel:2023rsl}. Even outside of the GW context, orbit averaging to obtain a secular approximation is well known to fail in three-body interactions~\cite{Antonini:2015zsa}, and more robust analysis methods have been developed to properly handle oscillatory orbital perturbations~\cite{NIT, krylov1949introduction}, such as the perihelion precession of Mercury~\cite{smith1985singular}.

For an accurate description of binary dynamics, especially from dynamical captures, a closer look is needed.
The equations of motion have to be defined locally, within the near-zone ---a region enclosing the source and much smaller than the typical wavelength of the radiation emitted \cite{gravitation,poisson}. There, the motion of the BHs is governed not only by their mutual Newtonian gravitational attraction, but also by dissipative terms accounting for GW emission.
This ``local'' approach underpins the post-Newtonian (PN) formalism \cite{1965ApJ...142.1488C, 1969ApJ...158...55C, poisson}. 

Translating the equations of motion into orbital evolution first requires a precise definition of the orbit itself~\cite{2019CQGra..36b5004L}. In general relativity (GR), there is no unique way to define an orbit, leading to two main approaches: (i) the Lagrangian planetary or osculating formalism, introduced by Lincoln and Will~\cite{1990PhRvD..42.1123L}, which relies on an intuitive Newtonian parametrization of the orbit, but with the orbital elements promoted to be functions of time; and (ii) the quasi-Keplerian formalism, developed by Damour and Deruelle~\cite{1985AIHPA..43..107D}, which employs a more accurate PN parametrization. %
However, the latter introduces complexities like multiple eccentricity parameters, complicating the interpretation of the orbital elements, while the former requires caution in interpreting osculating quantities physically~\cite{2019CQGra..36aLT01L, Will:2019lfe}.

Both approaches share a fundamental issue: they rely on instantaneous, local definitions of radiation-reaction terms, which are inherently gauge-dependent~\cite{Schaefer:1983hv, Thorne:1980ru, Burke:1970wx}. 
 Gauge dependencies are a constant feature in GR and represent the invariance of the theory under coordinate transformations~\cite{gravitation}. Not all components of the tensorial quantities characteristic of GR are meaningful. Certain gauges need to be chosen to remove spurious degrees of freedom and extract the physical content. A common example is the use of the Lorenz and Transverse-Traceless (TT) gauges in linearized theory of gravity to extract GW potentials \cite{maggiore}. Although all gauges have a common origin, they are not all equivalent.

In this work, we focus %
 solely on radiation-reaction gauges, which first enter the PN equations of motion at 2.5PN order.
We define radiation-reaction gauges as those that arise when we artificially partition regions of spacetime to simplify the treatment of the two-body problem. For instance, the PN formalism is valid only in the near zone. At larger separations (which include $\infty$), this expansion diverges and a different formalism has to be employed \cite{maggiore}.
To bridge these regimes, the two approaches must be connected in a region where their validity overlaps. This procedure, known as asymptotic matching \cite{1995PhRvD..52.6882I}, directly relies on defining such a boundary. However, this boundary is not a distinct region of spacetime but rather a conceptual limit imposed by our tools. Consequently, this locally (and arbitrarily) defined distance from the source can vary depending on the reference frame, introducing ambiguity that manifests as radiation-reaction gauge parameters \cite{1995PhRvD..52.6882I, maggiore, poisson}. These parameters enter the definition of the energy and angular momentum of the system, and the freedom in choosing them renders the evolution of the orbital elements ambiguous, complicating the extraction of physically meaningful insights. %

\subsection{Executive summary}

How, then, do eccentric BH binaries unambiguously evolve under GW emission, and across all relevant timescales? The orbit-averaged equations by Peters  \cite{1964PhRv..136.1224P} provide an approximate and limited description. While osculating methods yield more precise equations of motion, they lead to orbital evolutions that are intrinsically ambiguous.
In this work, we derive a novel, gauge-free, non-adiabatic set of evolutionary equations for the orbital elements of eccentric BH binaries, grounded in the osculating description of the orbit. %
Specifically, by ``gauge-free'' we mean that the equations we present are explicitly independent of the 2.5PN radiation-reaction gauge parameters. Further, in the remainder of this work, the term ``gauge parameters'' refers exclusively to the 2.5PN radiation reaction parameters.

We achieve this by redefining the orbit using the following mapping:
\begin{align}\label{mapping}
y&=\bar{y}+\frac{1}{c^5} \delta \bar{y},
\end{align}
where $y$ represents any of the orbital parameters including eccentricity $e$, semi-latus rectum $p$ [which is related to the semi-major axis $a$  through $p=a(1-e^2)$], true anomaly $f$,  %
longitude of the pericenter $\omega$, as well as  the time variable $t$.
The $\delta  \bar{y}$ functions are defined to ensure that the evolution equations for $\bar{y}$ are free from radiation-reaction gauge dependencies. We shall refer to the barred quantities $\bar{y}$ as \textit{characteristic parameters}, as these are not necessarily the geometric (or orbital) parameters describing the gauge-dependent orbital trajectory. 

Radiation-reaction gauge parameters enter the dissipative terms of the osculating equations \cite{1995PhRvD..52.6882I}, which are proportional to $c^{-5}$ and would vanish under orbit averaging. However, our reparametrization achieves gauge freedom without relying on this approximation.
We find the  $\delta \bar{y}$ functions by requiring that the expressions for the binary’s energy and angular momentum recover their Newtonian forms when expressed in terms of the new characteristic variables $\bar{y}$ and applying near-identity transformations~\cite{2024CQGra..41v5002L}.

The resulting evolutionary equations of  the characteristic parameters: %
\allowdisplaybreaks[4]
\begin{align} \label{dpbdtb}
  \frac{\dd \bar{p}}{\dd \bar{t}} &= -\frac{8}{5}\frac{\eta}{c^5}\frac{M^3}{\bar{p}^{\,3}}(1+\bar{e} \cos \bar{f})^3 \left[ 8 + 12 \bar{e} \cos \bar{f} + \right. \nonumber \\
&\quad \left. \bar{e}^{\, 2} \left(1 + 3 \cos 2\bar{f}\right) \right],  \\ \label{debdtb}
  \frac{\dd \bar{e}}{\dd \bar{t}} &= -\frac{2}{15}\frac{\eta}{c^5}\frac{M^3}{\bar{p}^{\, 4}}(1+\bar{e} \cos \bar{f})^3 \left[ 72 \cos \bar{f} + \right. \nonumber \\
&\quad \bar{e} \left(116 + 52 \cos 2 \bar{f} \right) + \bar{e}^{\,2} \left(109 \cos \bar{f} + 11 \cos 3 \bar{f} \right) \nonumber \\
&\quad \left. + \bar{e}^{\,3} \left(6 + 18 \cos 2 \bar{f} \right) \right], \\  \label{dfbdtb}
   \frac{\dd \bar{f}}{\dd \bar{t}} &= \frac{M^{1/2}}{\bar{p}^{\, 3/2}} \left(1 + \bar{e} \cos \bar{f} \right)^2, \\ \label{dwbdtb}
  \frac{\dd \bar{\omega}}{\dd \bar{t}} &= 0.
\end{align}
This set of equations describes the evolution of eccentric binaries in a manner consistent with the established osculating formalism while eliminating unphysical artifacts arising from radiation-reaction gauge parameters. %

We leverage our new equations to test for violations of the adiabatic assumptions in the formalism of Ref.~\cite{1964PhRv..136.1224P},
assessing the regions of the $(\bar{p}, \bar{e})$ parameter space where it can or cannot be employed. We confirm that the breakdown of the adiabatic formalism is connected to the timescales on which the binary evolution takes place. However, we show that knowledge of the initial conditions alone is insufficient to establish the validity of Peters' equations. Instead, one must examine the behavior of the timescales at the first pericenter passage.

This paper is organized as follows. In Sec.~\ref{radiation reaction}, we provide a brief introduction to balance laws, introduce their application to describe the dynamics of BH binaries, and present both the Peters' and the osculating-orbit formalisms. In Sec.~\ref{fixing2.5}, we present the derivation of Eqs.~\eqref{dpbdtb}–\eqref{dwbdtb}. In Sec.~\ref{evolution}, we compare the evolution of BH binaries using different approaches, investigate the breakdown of the orbit-average approximation, and study the impact of gauge parameters. Finally, in Sec.~\ref{conclusion}, we summarize our findings and outline future prospects.

\section{Radiation reaction}\label{radiation reaction}

\subsection{Conservation laws}\label{conservation laws}

In Newtonian mechanics, knowing all forces in a system defines its equations of motion. However, in complex systems, this information is hard to obtain. Here, conservation laws provide key insights.
Conservation laws are linked to the system's stress-energy tensor $\tensor{T}{^\alpha^\beta}$ and are expressed through its divergence, which vanishes when all relevant contributions are included:  
\begin{align} \label{commaT}
\tensor{T}{^\alpha^\beta}{,_\beta} = 0,
\end{align}  
where the comma denotes a partial derivative.  As a practical example, consider a perfect fluid in special relativity. By defining its stress-energy tensor and substituting it into Eq.~\eqref{commaT}, one obtains the Euler equations of motion  (see Ref.~\cite{poisson} for a complete derivation).  

Equation~\eqref{commaT} represents a conservation law in flat spacetime, as it enables the definition of globally conserved quantities, such as total energy and angular momentum. This is due to the presence of partial derivatives, which allow differential identities to be transformed into integral ones. Specifically, total energy and angular momentum can be determined by integrating Eq.~\eqref{commaT} over the entire three-dimensional space enclosing the source.  

This property of Eq.~\eqref{commaT} is lost in linearized gravity, where $g_{\mu \nu} \sim \eta_{\mu \nu} + \epsilon h_{\mu \nu} + {\cal{O}}(\epsilon^{2})$ with $\eta_{\mu \nu}$ the flat Minkowski metric, $h_{\mu \nu}$ the linearized metric perturbation, and $\epsilon$ an order-keeping parameter. In GR, ordinary derivatives are replaced by covariant derivatives, and commas are substituted with semicolons. Behind this ``extra dot''  lies the nearly fifty-year-long debate on the existence of GWs and the definition of the energy and momentum of the gravitational field \cite{1962gait.book.....W, gravitation, poisson, 2007tste.book.....K}. In GR, Eq.~\eqref{commaT} becomes
\begin{align} \label{semicolumnT}
\tensor{T}{^\alpha^\beta}{;_\beta} = \left(\sqrt{-g} \tensor{T}{^\alpha^\beta} \right)_{,\beta} + \tensor{\Gamma}{^\alpha_\beta_\gamma} \left(\sqrt{-g} \tensor{T}{^\beta^\gamma}\right) = 0,
\end{align}  
where %
$g$ is the metric determinant, and $\tensor{\Gamma}{^\alpha_\beta_\gamma}$ are the Christoffel symbols, which are of ${\cal{O}}(\epsilon)$.
The first term resembles the left-hand side of Eq.~\eqref{commaT}, while the second term indicates why Eq.~\eqref{semicolumnT} no longer describes the conservation of energy and angular momentum. Specifically, this term prevents the definition of globally conserved quantities solely from $\tensor{T}{^\alpha^\beta}$ as one can always define a coordinate system where $\Gamma^{\alpha}_{\beta \gamma} = 0$~\cite{gravitation}.

The strategy is to consider a quantity $\tensor{t}{^\alpha^\beta}$ that represents the distribution of gravitational energy and momentum and is conserved alongside the matter stress-energy tensor. This leads to a conservation law analogous to Eq.~\eqref{commaT}:
\begin{align} \label{commaT+t}
\left(\tensor{T}{^\alpha^\beta} +\tensor{t}{^\alpha^\beta}\right)_{,\beta}= 0.
\end{align}
Various definitions for $\tensor{t}{^\alpha^\beta}$ have been proposed in the literature (see Refs. \cite{2018CQGra..35e5011S, 2012PhRvD..85j4012S} for a comprehensive review). In this work, we adopt the formulation first put forward by Landau \& Lifshitz \cite{landau}, which arises from an exact reformulation of Einstein's equations \cite{landau, maggiore, poisson} and leads to the popular pseudo-tensor $\tensor{t}{^\alpha^\beta}$. %

One might be tempted to interpret this quantity %
as the energy and momentum of the gravitational field. However, caution is required here. Due to its non-tensorial nature, the definition of $\tensor{t}{^\alpha^\beta}$ is not unique; its value can change under coordinate transformations, and it can even be made to vanish in some frames \cite{landau, 1968PhRv..166.1263I,1968PhRv..166.1272I,graviataion, maggiore, poisson}. This is not surprising and aligns with a the equivalence principle in GR, which asserts that local gravitational effects can be nullified through an appropriate choice of coordinates.

For this reason, such pseudotensors cannot be used to infer the local back-reaction of a system due to gravitational radiation and cannot be directly used to define the equations of motion. However, Eq.~\eqref{commaT+t} allows for the definition of globally conserved quantities and, through further algebraic manipulation, provides definitions for the fluxes of energy and angular momentum carried to infinity by gravitational radiation \cite{1980RvMP...52..299T, poisson}.    %

It is important to note that Eqs.~\eqref{commaT} and \eqref{commaT+t}, although both representing conservation laws, are conceptually distinct \cite{poisson}. The first arises from the local conservation of energy and momentum, making it a fundamental principle that is entirely independent of Einstein's equations in flat spacetime. In contrast, the second directly relies on Einstein's equations, as they are explicitly used in the definition of the stress-energy pseudotensor.%
 
By integrating Eq.~\eqref{commaT+t} over a surface $S_j=R^2N_j$, where $R=|\vec{x}|$ is the distance from the source and $\vec{N}=\vec{x}/R$ is a unit vector, and assuming no intersection with matter while taking $R \to \infty$, the fluxes of energy and angular momentum transported by GW are defined, respectively, as
 \begin{align} \label{fluxes1}
\mathcal{P}& = c \oint_{\infty} (-g)\tensor{t}{ ^0^k} \dd S_k ,\\
\label{fluxes2}
 \tensor{\mathcal{F}}{^j^k}&= \oint_{\infty}\left[ x^j(-g)\tensor{t}{ ^k^n}-x^k(-g) t^{jn}\right] \dd S_n.
\end{align}

To ensure a clear physical meaning, the integrals in Eqs.~\eqref{fluxes1} and \eqref{fluxes2} must be evaluated in asymptotically Minkowskian coordinates. If this condition is not met, i.e. the spacetime is not flat at infinity, Eqs.~\eqref{fluxes1} and \eqref{fluxes2}, along with all quantities related to them, including total energy and angular momentum, become meaningless~\cite{gravitation}. Therefore, we are interested only in the definition of $\tensor{t}{^\alpha^\beta}$ in this regime, where we can also assume that the characteristic wavelength of GWs is much smaller than the background curvature scale. In other words, we must operate in the short-wavelength limit \cite{1968PhRv..166.1263I,1968PhRv..166.1272I,gravitation, poisson, maggiore}. 

By exploiting the definition of the Landau-Lifshitz pseudotensor in this regime and applying the Brill-Hartle averaging procedure~\cite{1964PhRv..135..271B}, along with imposing the Lorentz and TT gauge, one finally obtains a physically meaningful expression for $\tensor{t}{^\alpha^\beta}$. This averaging method, first introduced in Ref. \cite{1964PhRv..135..271B} and later applied in the GW context for the first time in Refs. \cite{1968PhRv..166.1263I,1968PhRv..166.1272I}, is a necessary step as it enables a clear identification of the effects of GWs on the background spacetime. Consequently, it allows a practical definition of the energy and angular momentum content of GWs only \cite{maggiore}. 
The resulting Landau-Lifshitz pseudotensor depends entirely on the derivatives of the usual GW potentials expressed in the TT gauge
\begin{align} \label{tLL}
 t^{\alpha \beta} &= \frac{c^4}{32 \pi}   \llangle[\Bigg] h^{\rm TT\,,\alpha} _{\gamma \delta}
 \,\,h_{\rm TT} ^{ \gamma \delta \, , \beta}
\rrangle[\Bigg],
\end{align}
where $\llangle  \cdot \rrangle$ is the Brill-Hartle average~\cite{1964PhRv..135..271B,1968PhRv..166.1272I}. We stress that such expression is invariant under any gauge transformation (see Ref. \cite{1968PhRv..166.1272I, maggiore, poisson,2012PhRvD..85j4012S} for a formal proof).

Plugging Eq.~\eqref{tLL} into Eqs.~\eqref{fluxes1} and \eqref{fluxes2} returns the fluxes expressed solely in terms of the gravitational potentials while preserving gauge-invariance. 
 Specifically, defining the gravitational potentials in terms of the mass quadrupole tensor $I^{jk}$ of the system leads to the famous quadrupole formula  \cite{poisson,maggiore}%
\begin{align} \label{fluxesnogauge1}
\mathcal{P}& = \frac{1}{5 c^5} \dddot{I}^{\langle jk\rangle} \dddot{I}^{\langle jk\rangle}+ \mathcal{O}(c^{-7}),\\
\label{fluxesnogauge2}
\mathcal{F}^{jk}& = \frac{2}{5 c^5} \left(\ddot{I}^{\langle jp\rangle} \dddot{I}^{\langle kp\rangle}–\ddot{I}^{\langle kp\rangle} \dddot{I}^{\langle jp\rangle}\right) + \mathcal{O}(c^{-7}),
\end{align}
where the overdots denote time derivatives and the angular brackets indicate the symmetric trace-free (STF) operation $n^{\langle jk\rangle}=n^jn^k-\frac{1}{3 }\delta ^{jk}$. %
With Eqs.~\eqref{fluxesnogauge1} and \eqref{fluxesnogauge2} established, the motion is described by %
 the related balance equations
\begin{align}
\label{Econservationlaws}
\frac{\dd \mathcal{E}}{\dd t} &= -\mathcal{P}, \\
\label{Lconservationlaws}
\frac{\dd \tensor{\mathcal{L}}{^j^k}}{\dd t} &= -\tensor{\mathcal{F}}{^j^k},
\end{align}
where $\mathcal{E}$ represents the total energy of the system, and $\tensor{\mathcal{L}}{^j^k}$ denotes the total angular momentum.

\subsection{Two-body dynamics}  \label{sec:rr}
Conservation laws, and consequently balance equations, have been extensively employed to describe the dynamics of two-body systems with varying degrees of accuracy. Equations~\eqref{Econservationlaws} and \eqref{Lconservationlaws} require knowledge of the energy, angular momentum, and fluxes. Expressions for these quantities have been central to the development of PN theory, and are now determined up to 4.5 PN order~\cite{1995PhRvD..52.6882I,1997PhRvD..55.6030G,1997PhRvD..55..714B,2009PhRvD..80l4018A,2002LRR.....5....3B}

Before analyzing the effects of radiation reaction on the dynamics of binary BHs, let us first establish the Newtonian framework for the problem. %
Consider a system composed of two bodies with masses  $m_1$ and $m_2$ orbiting each other. This is equivalently described as a single particle of mass $ \mu=m_1m_2/(m_1+m_2)$ moving around a larger object of mass $M=m_1+m_2$ . The effective particle experiences an acceleration
\begin{align}
\vec{a}_{\rm N}=-\frac{M}{r^2} \vec{n},
\end{align}
where $\vec{r}$ is the relative separation between the two bodies, $r=|\vec{r}|$ and $\vec{n}=\vec{r}/r$. 
The Newtonian energy $E_{\rm N}$ and angular momentum  $\vec{L}_{\rm N}$ of the particle are given by
\begin{align}\label{EN}
E_{\rm N}&=\frac{1}{2} \mu v^2- \frac{ \mu M }{r},\\
\label{LN}
\vec{L}_{\rm N}&=\mu \vec{r} \times \vec{v},
\end{align}
where $\vec{v}$ is the relative velocity of the particle.
Considering the radial and azimuthal coordinates $ r$ and $ \phi $, we can define a basis formed by the vectors $\vec{n} = [\cos \phi, \sin \phi, 0] $, $ \bm{\lambda} = [-\sin \phi, \cos \phi, 0] $, and $ \vec{e}_z=[0,0,1]$, which is normal to the orbital plane and parallel to $\vec{L}_{\rm N} $. One has
\begin{align}\label{def1}
\vec{r} &= r \vec{n}, \\ 
\label{def2}
\vec{v} &= \dot{r} \vec{n} + r \dot{\phi} \bm{\lambda}, \\
\label{def3}
\vec{a}_{\rm N} &= \left( \ddot{r} - r \dot{\phi}^2 \right) \vec{n} + \left( 2 \dot{r} \dot{\phi} + r \ddot{\phi} \right) \bm{\lambda}.
\end{align}
The Newtonian conservation of energy and angular momentum relates the terms appearing in Eqs.~\eqref{def1}, \eqref{def2}, and~\eqref{def3} to the orbital elements characterizing the binary%
\begin{align}\label{spatial_kepler1} 
r &= \frac{p_{\rm K}}{1 + e_{\rm K} \cos f_{\rm K}}, \\
\label{spatial_kepler2}
\dot{r} &= \sqrt{\frac{M}{p_{\rm K} }} e \sin f_{\rm K}, \\
\label{spatial_kepler3}
\dot{\phi} &= \sqrt{\frac{M}{p_{\rm K} ^3}}\left[1 + e_{\rm K}  \cos f_{\rm K} \right]^2, \\
v &= \sqrt{\frac{M}{p_{\rm K} } \left(1 + 2 e \cos f_{\rm K}  + e^2 \right)} \label{spatial_kepler4},
\end{align}
where we added a subscript $\rm K$ (``Keplerian''), to remind us that these quantities are constants on Keplerian orbits.

The effects of GW emission have to be accounted for on two distinct scales: globally and locally \cite{poisson}. On the global scale, the total time derivative of the energy and angular momentum no longer vanishes; instead, due to GW emission, it equals the fluxes given in Eqs.~\eqref{fluxesnogauge1} and \eqref{fluxesnogauge2}, which at leading order are expressed as~\cite{1993PhRvL..70..113I,1995PhRvD..52.6882I} %
\begin{align} \label{Pfluxesbinary}
\mathcal{P}& =- \frac{8}{5 c^5} \eta^2 \frac{M^2}{r^4} \left( 4 v^2 -\frac{11}{3} \dot{r}^2 \right),\\
\label{Ffluxesbinary}
\mathcal{F}^{jk}& =- \frac{8}{5c^5} \eta \vec{L}_{\rm N} \frac{M^2}{r^3} \left( 2 v^2+2 \frac{M}{r}- 3 \dot{r}^2 \right).
\end{align} 

Locally, i.e. in the near zone, the system dynamics have to react to what is happening globally i.e., in the far wave zone. In other words, equations of motion %
 must include dissipative terms that account for the radiation of GW to infinity \cite{poisson}.
Such dissipative terms are proportional to $c^{-5}$ and, at the leading order, result in an additional acceleration term on the particle
\begin{align}
\vec{a} &= -\frac{M \vec{n}}{r^3} \left[1 + \mathcal{O}(c^{-5})  \right] = \vec{a}_{\rm N}  +\vec{a}_{\rm 2.5PN}\,.\end{align}
In this expression, conservative terms, characterized by even powers of $c^{-1}$, and higher order corrections for the dissipative term have been omitted (cf.~Refs.~\cite{1995PhRvD..52.6882I, poisson}).
The term $\vec{a}_{\rm 2.5PN}$ models the radiation-reaction contribution to the acceleration, and can only depend on $v^2$, $m/r$, and $\dot{r}^2$ \cite{1990PhRvD..42.1123L,1993PhRvL..70..113I,1995PhRvD..52.6882I}. In particular, one has
\begin{align} \label{arrgeneral} 
\vec{a}_{\rm 2.5PN} = -\frac{8}{5} \frac{\eta M^2}{r^3} \left( - A_{\rm 2.5PN} \,\dot{r} \vec{n} + B_{\rm 2.5PN}\, \vec{v} \right), \end{align} 
where 
\begin{align} \label{Acoeff} A_{\rm 2.5PN} &= a_1 v^2 + a_2 \frac{M}{r} + a_3 \dot{r}^2, \\
 \label{Bcoeff} B_{\rm 2.5PN} &= b_1 v^2 + b_2 \frac{M}{r} + b_3 \dot{r}^2, \end{align} 
and the $a_{i}$'s and $b_{i}$'s are coefficients that still need to be determined.
To this end, Ref.~\cite{1995PhRvD..52.6882I} employs asymptotic matching techniques to connect the near-zone solution for the dynamics with the far-zone wave fluxes. Here is where radiation-reaction gauges come into play.

In practice, one must first define $\mathcal{E}$ and $\mathcal{L}$ to include contributions at  2.5PN order
\begin{align} \label{Eosculating} 
\mathcal{E} &= E_{\rm N} + E_{\rm 2.5 PN}, \\
\label{Losculating} \bm{ \mathcal{L} } &= \boldsymbol{L}_{\rm N} + \vec{L}_{\rm 2.5 PN}. \end{align} 
Following the same reasoning used for writing down the generic definition of $\vec{a}_{\rm 2.5PN}$, we have 
\begin{align} \label{E2.5} 
E_{\rm 2.5 PN} &= -\frac{8}{5} \frac{\eta M^2}{c^5 r^2} \dot{r} \left( p_1 v^2 + p_2 \frac{M}{r} + p_3 \dot{r}^2 \right), \\ 
\label{L2.5} 
\vec{L}_{\rm 2.5 PN} &= \frac{8}{5} \vec{L}_{\rm N} \frac{\eta M}{c^5 r} \dot{r} \left( q_1 v^2 + q_2 \frac{M}{r} + q_3 \dot{r}^2 \right), 
\end{align} 
where $p_{i}$ and $q_{i}$ are again coefficients to be determined, 
and
\begin{align} 
\label{Edev} 
\frac{\dd \mathcal{E}}{\dd t} &= \vec{v} \cdot \vec{a}_{\rm 2.5PN} + \frac{\dd E_{\rm 2.5 PN}}{\dd t}\,, \\ 
\label{Ldev} 
\frac{\dd \bm{\mathcal{L}}}{\dd t} &= \vec{r} \times \vec{a}_{\rm 2.5PN} + \frac{\dd \vec{L}_{\rm 2.5 PN}}{\dd t} .
\end{align}
Finally, one must match these expressions with the fluxes evaluated at the same PN order \cite{1993PhRvL..70..113I,1995PhRvD..52.6882I}, i.e. one must equate the right-hand-side of Eqs.~\eqref{Edev}-\eqref{Ldev} with Eqs.~\eqref{Pfluxesbinary}-\eqref{Ffluxesbinary}. Since this work considers terms up to 2.5PN order, the fluxes defined in Eqs.~\eqref{Pfluxesbinary} and \eqref{Ffluxesbinary} can be used for this purpose.

There are twelve undetermined coefficients in Eqs.~\eqref{Acoeff}, \eqref{Bcoeff}, \eqref{E2.5}, and \eqref{L2.5}. However, comparing the terms in Eqs.~\eqref{Edev} and \eqref{Ldev} with those in Eqs.~\eqref{Pfluxesbinary} and \eqref{Ffluxesbinary} specifies only ten of these coefficients. This leaves two degrees of freedom, represented by the arbitrary radiation-reaction gauge parameters  $\alpha$ and $\beta$ \cite{1993PhRvL..70..113I,1995PhRvD..52.6882I}. We report their resulting expression for such parameters in Table \ref{tab:coeff}.
\begin{table}[b]\label{tab:coeff}
\centering
\renewcommand{\arraystretch}{1.5} 
\begin{tabular}{c|ccc}
 & $i=0$ & $i=1$ & $i=2$ \\
\hline
$a_{i}$ & $3 + 3\beta$ & $\frac{23}{3} + 2\alpha - 3\beta$ & $-5 \beta$ \\
$b_{i}$ & $2 + \alpha$ & $2 - \alpha$ & $-3 - 3\alpha$ \\
$p_{i}$ & $-(\alpha + 2)$ & $0$ & $\beta$ \\
$q_{i}$ & $0$ & $\alpha$ & $0$ \\
\end{tabular}
\caption{Coefficients of Eqs.~\eqref{Acoeff}, \eqref{Bcoeff}, \eqref{E2.5} and \eqref{L2.5} determined by matching Eqs.~\eqref{Edev} and \eqref{Ldev} with the fluxes of Eqs.~(\ref{Pfluxesbinary}) and (\ref{Ffluxesbinary}). See Refs. \cite{1993PhRvL..70..113I,1995PhRvD..52.6882I}. }
\end{table}
As noted in Refs.~\cite{1995PhRvD..52.6882I, poisson}, the presence of  $\alpha$ and $\beta$ reflects how the relative orbital separation vector $\vec{r}$, explicitly present in Eqs.~\eqref{Edev} and \eqref{Ldev}, depend on the choice of coordinate frame. Although $\vec{r}$ is not a coordinate itself, its behavior is influenced by the adopted reference frame. %

Each choice of $\alpha$ and $\beta$ corresponds to a specific coordinate system. The harmonic coordinate system, commonly used in PN derivations \cite{maggiore, poisson,1992MNRAS.254..146J}, is defined by ($\alpha = -1$, $\beta = 0$) \cite{ 1995PhRvD..52.6882I,DAMOUR198181}. However, this framework introduces unnecessary complexity in both the metric and the radiation reaction force \cite{1983NCimL..36..105S, poisson}. As demonstrated in Ref. \cite{poisson}, alternative combinations of these radiation-reaction gauges can significantly simplify some of the metric components and, consequently, the radiation reaction potentials. Notable examples include the Burke-Thorne gauge ($\alpha = 4$, $\beta = 5$) \cite{1969ApJ...158..997T,1971JMP....12..401B,gravitation,1995PhRvD..52.6882I} and the Sch\"{a}fer gauge ($\alpha = 5/3$, $\beta = 3$) \cite{1983NCimL..36..105S}.
In the following, we compare our findings against equations of motions obtained with these three sets of gauge parameters.

\subsection{ Lagrangian planetary equations}\label{lagrangian}

Once the expression for $\vec{a}_{\rm 2.5PN}$ is determined, the method of osculating orbits provides a suitable framework to derive the evolution equations for the orbital elements; these are known as either Lagrangian planetary equations or osculating equations. %
The broad strategy is to make use of Newtonian description of the dynamics (Sec.~\ref{sec:rr}) and solve Eq.~\eqref{arrgeneral} by promoting the quantities introduced in Eqs.~\eqref{spatial_kepler1}–\eqref{spatial_kepler4} to functions of time, treating $\vec{a}_{\rm 2.5PN}$ as a perturbation to the Newtonian two-body problem \cite{poisson, 2004PhRvD..70f4028D}. Specifically, it assumes that, at any given time $t_{i}$, the system can be described by a Newtonian configuration characterized by a set of orbital elements

\begin{align} 
\mu^{\sigma}(t_{i}) = [p(t_{i}), e(t_{i}), f(t_{i}), \omega(t_{i})].
\end{align} 
Here, we drop the subscript K on these quantities because they are no longer constants of motion. Due to the perturbation, these elements differ from those describing the system at another time $t_{\rm j}$ (with $i \neq j$).
Practically, one needs to solve the following set of equations to recover the evolution of the orbital elements
\begin{align}
\frac{\partial \vec{r}}{\partial \mu^{\sigma}}\frac{\dd \mu^{\sigma}}{\dd t} &=0, \\
\frac{\partial \vec{v}}{\partial \mu^{\sigma}}\frac{\dd \mu^{\sigma}}{\dd t} &= \vec{a}_{\rm 2.5PN},
\end{align}
where $\vec{r}$ and  $\vec{v}$  are defined %
 as in Eqs.~\eqref{def1} and \eqref{def2}. To this end, it is useful to express $\vec{a}_{\rm 2.5PN}$ in the basis introduced in Sec.~\ref{sec:rr}
\begin{align}
 \vec{a}_{\rm 2.5PN}&=\mathcal{R} \vec{n}+\mathcal{S} \bm{\lambda},
 \\
 \label{R}
 \mathcal{R}&=\frac{8}{5c^5}\frac{ \eta M ^2}{r^3}  \dot{r} (A_{2.5} - B_{2.5} ),\\
  \label{S}
\mathcal{S}&=- \frac{8}{5 c^5} \left(\frac{M}{r}\right)^2 \eta  \dot{ \phi} B_{2.5} ,
\end{align}
where $A_{2.5}$ and  $B_{2.5}$ are defined respectively in Eqs. \eqref{Acoeff} and \eqref{Bcoeff} and their  $a_i$ and $b_i$ coefficients reported in Table \ref{tab:coeff}. The conversion between phase-space coordinates and orbital parameters is reported in Eqs. \eqref{spatial_kepler1}$-$\eqref{spatial_kepler4}. 
The resulting evolution equations for the orbital elements $\mu^{\sigma}$%
 are then
\begin{align}\label{dpdt}
\frac{\dd p}{\dd t} &= \sqrt{\frac{p^3}{M}}\frac{2}{(1+e\cos f)}\mathcal{S},\\ \label{dedt}
\frac{\dd e}{\dd t} &= \sqrt{\frac{p}{M}}\left[\sin f \,\mathcal{R}+\frac{2 \cos f+ e(1+ \cos^2 f)}{1+e\cos f}\mathcal{S}\right],\\\label{dfdt}
\frac{\dd f}{\dd t} &=\sqrt{\frac{M}{p^3}}  (1+e \cos f)^2+\frac{1}{e} \sqrt{\frac{p}{M}}\left[\cos f \, \mathcal{R}+ \right. \nonumber \\
&\quad \left. -\frac{2+e\cos f}{1+e\cos f} \sin f \,\mathcal{S} \right], \\ \label{dwdt}
\frac{\dd \omega}{\dd t} &= \frac{1}{e}\sqrt{\frac{p}{M}}\left[-\cos f \,\mathcal{R}+\frac{2 +e \cos f} {1+e\cos f}\sin f\, \mathcal{S}\right],
 \end{align}
where $\mathcal{S}$ and  $\mathcal{R}$ are defined in  Eqs. \eqref{R} and \eqref{S}. 
This set of equations captures the evolution of BH binary systems, including the orbital-timescale dynamics and the non-adiabatic emission of GWs. %
This formulation holds in both the parabolic and hyperbolic limits. %

\subsection{ Peters' equations} \label{peters}
The secular behavior of orbital elements is obtained by orbit averaging Eqs.~(\ref{dpdt})-(\ref{dwdt}). This yields
\begin{align} \label{eq:Peters1}
\Big\langle \frac{\dd p}{\dd t} \Big\rangle &= -\frac{64}{5 c^5} \frac{\mu M^2}{p^3} (1 - e^2)^{3/2} \left(1 + \frac{7}{8} e^2 \right), \\
\label{eq:Peters2}
\Big\langle \frac{\dd e}{\dd t} \Big\rangle &= -\frac{304}{15 c^5} \frac{\mu M^2}{p^4} e (1 - e^2)^{3/2} \left(1 + \frac{121}{304} e^2 \right),
\end{align}
which is the popular result by Peters~\cite{1964PhRv..136.1224P}. 
Here, $\langle \cdot \rangle$ denotes the orbit-averaging operation, which for a generic quantity $X$ is defined as
\begin{align}
\langle X \rangle = \frac{1}{\tau_{\rm orb}} \int_0^{\tau_{\rm orb}} X(t)  \dd t,
\end{align}
where
\begin{align}\label{torb}
\tau_{\rm orb} =\frac{ 2 \pi}{\sqrt{ M }} \left( \frac{p}{1-e^2}\right)^{3/2}
\end{align}
is the orbital period.
Specifically, Peters' approach relies on two key choices: (i) adopting the Newtonian definitions of energy and angular momentum, and (ii) averaging the resulting equations over the orbital period. Averaging removes the oscillatory variations occurring on the orbital timescale from the expressions for the energy, the angular momentum, and their corresponding fluxes.
Consequently, $\omega$ remains fixed during the inspiral, and the usage of Newtonian definitions is justified.

The orbit averaging approximation is valid whenever changes induced by radiation reaction are small over an orbital period. This requires the timescale \footnote{Note that this definition of the radiation reaction timescale is not unique, i.e. one could make other choices such as $\tau_{\rm rr} = e/|\dd e/\dd t|$ or $\tau_{\rm rr} = \omega_{\rm orb}/|\dd \omega_{\rm orb}/\dd t|$ where $\omega_{\rm orb}$ is the orbital frequency. We have checked that such alternative choices do not alter the final results of this work.}
\begin{align}\label{trr}
\tau_{\rm rr}=\frac{p}{|\dd p/\dd t|}
\end{align}
 on which GW emission modifies the orbital parameters to be much longer than the orbital period, i.e., $\tau_{\rm orb} \ll \tau_{\rm rr}$.
Additionally, the orbital period must be finite. As pointed out by some of us in Ref.~\cite{2023PhRvD.108l4055F}, Peters' equations predict that $ e\to 1$ for all eccentric binaries as $t\to - \infty$, and consequently $\tau_{\rm orb} \to \infty$. This invalidates the assumptions used to derive the equations themselves, leading to an incorrect evolution of the orbital elements. Finally, as noted in Ref. \cite{2023PhRvD.108l4055F}, the definition of the radiation-reaction timescale given in Eq. \eqref{trr} is not unique. In principle, it can be computed using different orbital elements and their respective time derivatives. While small differences may arise, there are no preferred parameters for defining $\tau_{\rm rr}$. We therefore choose to use the semi-latus rectum due to its straightforward interpretation.  %

\section{ Fixing the 2.5PN ambiguity}\label{fixing2.5}

\subsection{Why this paper}

The price to pay for adopting the osculating equations instead of Peters' is the presence of radiation-reaction gauge parameters. In Eqs.~\eqref{E2.5}, \eqref{L2.5}, and \eqref{arrgeneral}, the expressions for energy, angular momentum, and their evolution, and consequently the evolution of the orbital elements are inherently ambiguous, and this ambiguity is encoded in the free parameters $\alpha$ and $\beta$.

 As discussed in Sec.~\ref{sec:rr}, radiation-reaction gauges parameters enter the equations of motion at 2.5PN order.
The appearance of new terms in the definition of energy and angular momentum when including radiation reaction
 is not unique to gravity \cite{gravitation, poisson}. In electromagnetism, similar terms appear when considering the dynamics of charged bodies.  In that context ---and in complete analogy with binary BHs--- part of the system's energy and angular momentum are converted into radiation which propagates outward, leaving the system. When balance equations are used to describe the dynamics, one must consider the additional terms known as Schott energy and angular momentum \cite{griffiths,Schott, 2014shst.book..165G}.
 Specifically, such terms appear exactly as the 2.5 PN term in the right-hand-side of  Eqs.~\eqref{Edev}  \cite{poisson, 2016PhRvD..93l4010G, 2014shst.book..165G}. However, the interpretation of either electromagnetic waves or their energy content is not ambiguous. Here the analogy between electromagnetism and gravity cannot go any further than a mathematical resemblance.  Schott terms have a clear physical meaning: they represent bound field energy, a quantity that remains attached to the system and is reversibly exchanged between the electromagnetic field and the bodies \cite{poisson, 2014shst.book..165G}. 
 
In contrast, the 2.5PN terms in gravity directly result from dividing the spacetime into zones, which is required to connect the energy and angular momentum transported by GWs at infinity, where they can be clearly quantified and measured, to the dynamics near the source, where these quantities are not well-defined for the gravitational field.
The ambiguity in choosing the boundary where matching conditions are applied directly impacts the definition of the system's energy and angular momentum, making the physical interpretation of these quantities unclear.
 Resolving this issue is essential for accurately studying the dynamics of binary BHs. Specifically, our goal here is to derive osculating-like equations that do not depend on the parameters $\alpha$ and $\beta$.
 
As already discussed in Sec.~\ref{peters},  these gauge parameters can be eliminated by orbital averaging. Radiation-reaction terms appear in the equations of motion with odd powers of $c^{-1}$, as they are necessary to break the time-reversal invariance of the theory (radiation can only propagate outward). This also results in their disappearance when averaging is applied.
However, orbital averaging results in an approximate description of the evolution of binaries and cannot be applied across the entire parameter space. %
This limitation justifies the need for the present work.

Our strategy is to introduce new definitions of the orbital elements. These new characteristic variables, for which we provide a mapping to their Keplerian counterparts, allow us to derive a set of evolution equations where the radiation-reaction gauge parameters are completely eliminated. This procedure requires invoking near-identity transformations, which we now introduce.

\subsection{Near-identity transformations} \label{NIT}

Near-identity transformations (NITs) \cite{NIT} are a mathematical tool used to simplify the equations of motion of dynamical systems by removing undesired dependencies on certain parameters. For instance, in the study of extreme mass-ratio inspirals, NITs are often employed to eliminate oscillatory terms from the equations of motion, enabling more efficient and accurate integration~\cite{2018CQGra..35n4003V,2022CQGra..39n5004L}. 
Similar techniques have also been employed in Refs. \cite{2004PhRvD..70f4028D,2006PhRvD..73l4012K} to derive the evolutionary equations for the orbital elements using the quasi-Keplerian formalism, although retaining gauge parameters.
The core idea is to transform one's variables into a new set that retains the essential dynamics while suppressing unwanted dependencies in the equations of motion.
Consider a generic dynamical system described by $\vec{x} = (x_0, x_1, \ldots, x_n)$ and their evolution equations
  \begin{align}\label{eqmotion}
 \dot{x_{i}}&=f_{x_{i}}^0 (\vec{x})+\epsilon f_{x_{i}} (\vec{x}, k)+\mathcal{O}(\epsilon^2),
 \end{align}
where $i = 0, \ldots, n$, the overdot denotes differentiation with respect to a chosen variable (e.g., time). The functions $f_{x_i}^j$ depend on both the variables $\vec{x}$ as well as another parameter $k$, which we wish to eliminate. The goal of NITs is to find a transformation that maps these variables to a new set,  $\tilde{\vec{x}}=(\tilde{x}_{0}, \tilde{x}_{1},\cdots,\tilde{x}_{\rm n})$, such that
 \begin{align}\label{arrow}
   \tilde{\vec{x}}\xrightarrow {%
  \substack{\rm Inverse\\  \rm NIT} } \vec{x}+\mathcal{O}(\epsilon),
 \end{align}
whenever $\epsilon \ll 1$. In this transformed framework, the equations of motion for the new variables are independent of $k$
\begin{align}\label{eqmotionNIT}
 \dot{\tilde{x}}_{i}&=F_{\tilde{x}_{i}}^0(\tilde{\vec{x}})+\epsilon F_{\tilde{x}_{i}} (\tilde{\vec{x}})+\mathcal{O}(\epsilon^2).
 \end{align}
To achieve this, we assume the existence of a transformation satisfying the inverse condition  of Eq.~\eqref{arrow} in the form  \begin{align}\label{straigthNIT}
\vec{x}= \tilde{\vec{x}}+\epsilon X(\tilde{\vec{x}},k ),
 \end{align}
where $ X(\tilde{\vec{x}},k )$  is an unknown set of functions. By differentiating this expression, comparing it with the original equations of motion Eq. \eqref{eqmotion}, employing Eq.~\eqref{eqmotionNIT}, and, expanding in $\epsilon$, one gets
 \begin{align}\label{deltaNIT}
 \dot {X}(\tilde{\vec{x}},k )=f_{x_{i}}(\tilde{\vec{x}},k)-F_{\tilde{x}_{i}} (\tilde{\vec{x}}).
 \end{align}
Here, make use of the assumption that the difference between the original and transformed variables is always of $\mathcal{O}(\epsilon)$, implying $F_{\tilde{x}_{i}}^0 =f_{x_{i}}^0$.  For the same reason, in all terms proportional to $\epsilon$ we can freely switch between $\vec{x}$ and $\tilde{\vec{x}}$. 
By integrating Eq.~\eqref{deltaNIT} and choosing $F_{\tilde{x}_{i}}$ appropriately, one can ensure that iterating this procedure results in equations of motion for $\tilde{\vec{x}}$ that are free from the unwanted dependency on $k$. The strength of NITs lies in the flexibility to choose the unknown functions $F_{\tilde{x}_{i}} (\tilde{\vec{x}})$. 

We now apply this method to remove the dependency of the radiation-reaction gauge parameters $\alpha$ and $\beta$ from the osculating equations describing BH binaries on generic orbits.

\subsection{Our new set of evolutionary equations}\label{ourmethod}

\begin{figure}[t]
\begin{center}
\includegraphics[width=0.85\columnwidth]{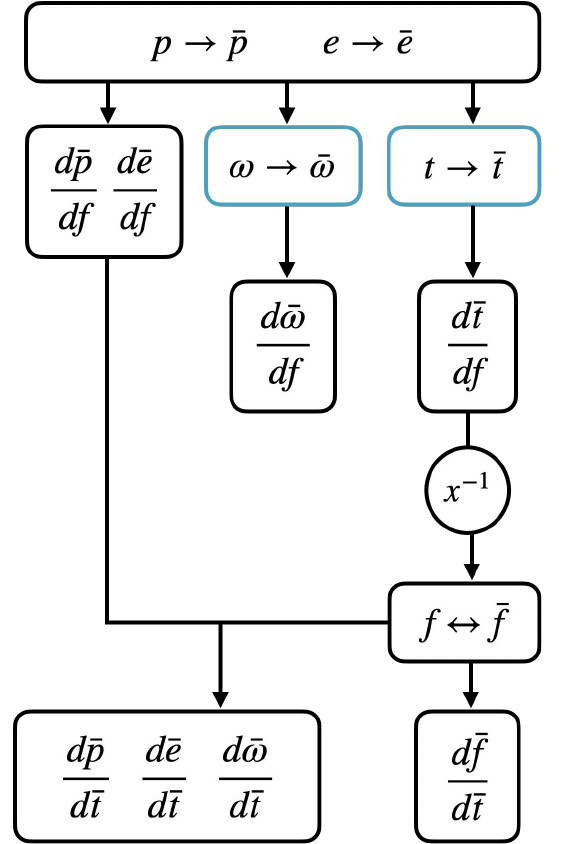} 
\caption{Derivation procedure for our new set of radiation-reaction, gauge-free evolutionary equations in terms of the characteristic parameters $\bar{p}$, $\bar{e}$, $\bar{\omega}$, $\bar{f}$.  Light blue boxes indicate steps where NITs are used. The circular box labeled $x^{-1}$ denotes where the inverse of the connected expression is taken. 
At any step, expansions in $c^{-1}$ are performed and only terms up to $\mathcal{O}(c^{-5})$ are kept.}
\label{diagram}
\end{center}
\end{figure} 

We map the osculating orbital parameters $(p, e, \omega, f, t)$ to a new set $(\bar{p},\bar{e},\bar{\omega},\bar{f}, \bar{t})$, and impose that their evolutionary equations are independent  of $\alpha$ and $\beta$. We proceed as follows:
\begin{itemize}
\item[i)] We use the definitions of energy and angular momentum from Eqs.~\eqref{Eosculating} and \eqref{Losculating} to define the mappings for $p(\bar{p},\bar{e},f)$ and $e(\bar{p},\bar{e},f)$.
\item[ii)] We them employ NIT (see Sec. \ref{NIT}) to handle the mappings for $\omega(\bar{p},\bar{e},f)$ and $t(\bar{p},\bar{e},f)$.
\item[iii)] The quantities $f$ and $\bar{f}$ can be exchanged freely at the PN order here considered. 
\end{itemize}
These mappings result in the evolutionary equations presented in Eqs.~\eqref{dpbdtb}–\eqref{dwbdtb} above. A schematic representation of our procedure is presented in Fig.~\ref{diagram}.

Before proceeding, let us comment on the ambiguities in this procedure. There is mathematical freedom in how one defines the mapping from the standard osculating quantities $\mu^{\sigma}$ to the new characteristic variables $\bar{\mu}^{\sigma}$. This is a result of the mathematical framework underpinning perturbation theory and is not related to gauge transformations in GR. Knowing this, how to construct physically meaningful mappings? First, we enforce that the secular evolution of the binary, that is Peters' equations in Eqs.~\eqref{eq:Peters1}–\eqref{eq:Peters2} at the PN order we are considering, should remain invariant. %
Second, we use the secular approximation as a guide to choosing the mappings. For example, upon orbit averaging one has $\langle d\omega/dt \rangle = 0$, despite the fact that the full osculating equation $d\omega/dt$ in Eq.~\eqref{dwdt} does not necessarily vanish. The secular equation informs us that the orbit should not precess under radiation reaction, at least at leading PN order. We can thus use this fact to define the new characteristic variable $\bar{\omega}$. Similarly, from Eq.~\eqref{Eosculating} and under orbit averaging, one has $\langle {\cal{E}} \rangle = E_{\rm N}$, which we use in part to obtain the characteristic variables $(\bar{p}, \bar{e})$.  The full details of our procedure are presented below.

As a first step, we map the semi-latus rectum and the eccentricity to their redefined counterparts, $\bar{p}$ and $\bar{e}$. This mapping is achieved by requiring the following relations:
\begin{align}  
\label{pbars}
p &= \bar{p} + \frac{1}{c^5} \delta \bar{p}\,(\bar{p}, \bar{e}, f, \alpha, \beta), \\ 
\label{ebars}
e &= \bar{e} + \frac{1}{c^5} \delta \bar{e}\,(\bar{p}, \bar{e},f, \alpha, \beta),
\end{align}
where $\delta \bar{p}$ and $\delta \bar{e}$ are generic unknown functions. To determine these functions, we substitute Eqs.~(\ref{pbars}) and \eqref{ebars} into the expressions for $\mathcal{E}$ and $\mathcal{L}$, given in Eqs.~\eqref{Eosculating} and \eqref{Losculating}, after expressing them in terms of the orbital elements using Eqs.~\eqref{spatial_kepler1}$-$\eqref{spatial_kepler4}.

Expanding the resulting expressions and retaining terms up to $\mathcal{O}(c^{-5})$, we impose the following conditions:
  \begin{align}\label{Ebar}
\mathcal{E} &= \frac{M^2 \eta}{2 \bar{p}} \left(1 - \bar{e}^2\right), \\ 
\label{Lbar}
\mathcal{L} &= M^{3/2} \eta \sqrt{\bar{p}}.
\end{align}
In doing this, i.e, requiring that the energy and angular momentum in the new barred variables are free from radiation-reaction gauge parameters, we find the following expressions for $\delta \bar{p}$ and $\delta \bar{e}$:
  \begin{align}
	\label{eq:delta-p}
 \delta \bar{p}&=-\frac{16} {5} \frac{M^{5/2}}{\bar{p}^{\,3/2}}\,\eta\,  (1+\bar{e} \cos f)^2 \,  \bar{e} \, \alpha\, \sin f,\\
	\label{eq:delta-e}
 \delta \bar{e}&=-\frac{4 }{5} \left(\frac{M}{\bar{p}}\right)^{5/2} \, \eta\,(1+\bar{e} \cos f)^2 \left\{4+4 \,\bar{e}\,(2+\alpha)\cos f+  \right. \nonumber \\
&\quad \left. +  \bar{e}^{\,2}[4+4\alpha-\beta+\beta \cos 2f]\right\} \sin f.
\end{align}
We specifically choose to employ $\mathcal{E}$ and  $\mathcal{L}$ to carry out this step, as the semi-latus rectum and eccentricity, along with their respective evolution equations, are directly derived from these quantities and their derivatives. Finally, we differentiate Eqs.~\eqref{pbars} and \eqref{ebars} with respect to $f$, solving for the derivatives of $ \bar{p}$, $ \bar{e}$. 

Since no other constants of motion, such as energy or angular momentum, are available to map the remaining orbital parameters, we employ NITs. Our goal is to eliminate the gauge parameters $\alpha$ and $\beta$ from $\dd \omega / \dd f$ and  $\dd t /\dd f$. As these parameters appear only in terms $\propto c^{-5}$, this approach is consistent with the method described in Sec.~\ref{NIT}.
We postulate the existence of the following NITs:
 \begin{align}
\label{wbars} \omega &= \bar{\omega} + \frac{1}{c^5} \delta \bar{\omega}(\bar{p}, \bar{e}, f, \alpha,\beta), \\
\label{tbars} t &= \bar{t} + \frac{1}{c^5} \delta \bar{t}(\bar{p}, \bar{e}, f, \alpha, \beta),
\end{align}
where $ \delta \bar{\omega}$ and $\delta \bar{t}$ are unknown functions designed to remove the dependency on the gauge parameters.
From Sec.~\ref{NIT}, we define:
 \begin{align}
\label{deltaw}
 \delta \bar{\omega}& =\int g(\bar{p}, \bar{e}, f, \alpha,\beta)-G(\bar{p}, \bar{e}, f) \dd f ,\\
 \label{deltat}
 \delta \bar{t}& =\int s(\bar{p}, \bar{e}, f, \alpha,\beta)-S(\bar{p}, \bar{e}, f) \dd f, \end{align}
where $g$  is the right-hand side of Eq.~\eqref{dwdt} divided by that of Eq.~\eqref{dfdt}. This function can be decomposed as $g=g_{\rm N}+c^{-5} g_{2.5PN}$. Using this definition, we set $G=g_{\rm N}$. Similarly,  $s$ is defined as the reciprocal of the right-hand side of Eq.~\eqref{dfdt}, rewritten as $s=s_{\rm N}+c^{-5} s_{2.5}$, from which we define $S=s_{\rm N}$.  In all cases, we apply the mappings of Eqs.~\eqref{ebars} and \eqref{pbars} and expand the resulting expressions in powers of $c^{-1}$, truncating them at $\mathcal{O}(c^{-5})$, obtaining the following expression for $\delta \bar{\omega}$ and $\delta \bar{t}$:
\begin{align}
\delta \bar{\omega}&=\frac{1}{180}\frac{M^{5/2} }{ \bar{p}^{5/2}}\frac{\eta}{ \bar{e}} \left \{2304 \cos \bar{f}+96 \,\bar{e}\, (23+3 \alpha) \cos2 \bar{f} \,+ \right. \nonumber \\
&\quad  +\bar{e}^{\,2} [12 \,(206-24 \alpha+18 \beta) \cos f+8\, (127+36 \alpha\,+\nonumber \\
&\quad+ 9 \beta) \cos3 \bar{f}]+\bar{e}^{\,3} [12\, (65+12 \beta) \cos 2 \bar{f}+3\, (59\,+\nonumber \\
&\quad
+\,24 \alpha+24 \beta )\cos4 \bar{f}\,]+\bar{e}^{\,4} [36 \,(6+\beta) \cos\bar{f}+6\, (12\,+\nonumber \\
&\quad  \left. +\,7 \beta) \cos 3 \bar{f} +18 \,\beta \cos 5 \bar{f}\,]\right\},\\
\delta \bar{t}&=\frac{2}{15} \frac{M^2}{  \bar{p}}\frac{ \eta}{\bar{e}^{\,2} }  \left\{12 \log(1+ \bar{e} \cos  \bar{f}\,)+84  \,\bar{e} \cos \bar{f}\,+  \right. \nonumber \\
&\quad
 +\bar{e}^{\,2} [(35+12 \alpha) \cos2  \bar{f}–12 \log(1+ \bar{e} \cos  \bar{f}\,)]\,+\nonumber \\
&\quad  \left. +\, \bar{e}^{\,3} [ (-12+12 \alpha \,  -\,9\, \beta) \cos \bar{f}+3\, \beta \cos 3  \bar{f})]\right\}.
\end{align}
By differentiating Eqs.~\eqref{wbars} and \eqref{tbars} with respect to $f$, using Eqs.~\eqref{dwdt} and \eqref{dfdt}, employing the new definition of $\bar p$ and $\bar e$ and, expanding the expressions in powers of $c^{-1}$ retaining the terms up to  $\mathcal{O}(c^{-5})$ we finally obtain $\dd \bar{\omega}/\dd f$ and $\dd \bar{t}/\dd f$.

To derive Eqs.~\eqref{dpbdtb}, \eqref{debdtb}, and \eqref{dwbdtb}, we firstly divide $\dd \bar{p}/\dd f$, $\dd \bar{e}/\dd f$, $\dd \bar{\omega}/\dd f$  by $\dd \bar{t}/\dd f$. Then, we redefine $f$ as
\begin{align}
\label{fbars} f &= \bar{f} + \mathcal{O}(c^{-5}). 
\end{align}
However, since $f$ always appears as the argument of trigonometric functions in all expressions, expanding the resulting equations in powers of  $c^{-1}$ and truncating them at $\mathcal{O}(c^{-5})$ ensures that $f$ is always interchangeable with~$\bar{f}$. %

It will not go unnoticed that Eqs.~\eqref{dwbdtb} and \eqref{dfbdtb} align with those typically employed in a Newtonian framework. This is because, following the approach of Refs.~\cite{poisson,1995PhRvD..52.6882I}, we do not include  1PN and 2PN terms in our derivation. GWs solely extract energy and angular momentum from the system, leading only to variations in the semi-latus rectum and eccentricity. Since our analysis is restricted to radiation reaction effects, phenomena such as periastron precession \cite{1988NCimB.101..127D, poisson} are not captured by these equations, consistently with Ref~\cite{2004PhRvD..70f4028D}, but could be straightforwardly added using the methods of, e.g.,   Refs.~\cite{1985AIHPA..43..107D, Mora:2003wt}.

Finally, we provide some further information on the gauge parameters in this formalism. First, it is straightforward to show, by direct comparison, that no choice of  $(\alpha, \beta)$ in Eqs.~\eqref{dpdt}–\eqref{dfdt} can reproduce Eqs.~\eqref{dpbdtb}–\eqref{dfbdtb}. Thus, the characteristic variables and their evolutionary equations are not merely a suitable gauge choice. Second, while we have removed the gauge parameters $(\alpha,\beta)$ from the evolution equations of the binary, they are, however, not completely removed from the two-body problem. The osculating formalism from which we started uses the mappings in Eqs.~\eqref{spatial_kepler1}–\eqref{spatial_kepler4} for the orbital trajectory. Inserting these into the mappings in Eq.~\eqref{pbars}–\eqref{ebars} and~\eqref{eq:delta-p}–\eqref{eq:delta-e} reveals that the gauge parameters now appear inside of these coordinate trajectories. This is unsurprising: the gauge parameters describe different coordinate systems, and the trajectory is a coordinate-dependent quantity. Further, gauge parameters naturally arise as a result of extending perturbation theory to higher orders. The benefit of the characteristic parameters that we have derived is that they satisfy dynamical equations that are independent of these coordinate effects, and thus, one can obtain an unambiguous picture of BH binary dynamics in PN theory from them. Lastly, consistent with the osculating formalism, our equations remain valid for arbitrary values of eccentricity. %

\begin{figure*}[p]
\begin{center}
\includegraphics[width=0.95\textwidth]{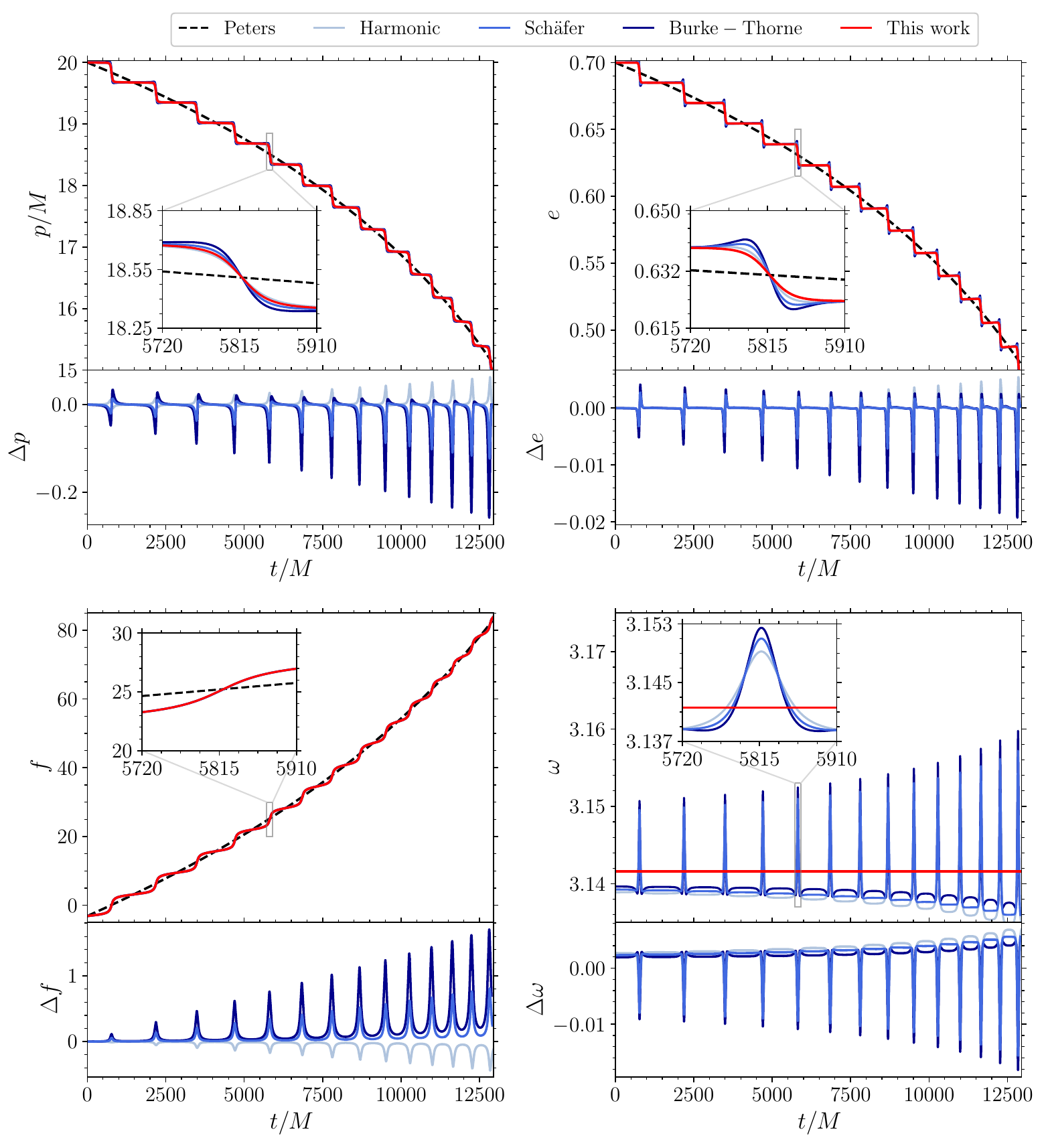} 
\caption{Evolution of the semi-latus rectum (top-left panel), eccentricity (top-right panel), true anomaly (bottom-left panel), and longitude of the pericenter (bottom-right panel) for a binary system with initial conditions $(\bar{p}_0,\bar{e}_0,\bar{f}_0,\bar{\omega}_0)=(20 M, 0.7, -\pi, \pi)$.
Binaries are evolved using different sets of equations: Peters' equations (dashed grey curve), Lagrangian planetary equations in the harmonic gauge (solid light blue curve), the Schäfer gauge (solid blue curve), and the Burke-Thorne gauge (solid dark blue curve), as well as the new set of equations presented in this paper (solid red curve). The bottom subpanels of each panel show differences between the evolution obtained using the Lagrangian planetary equations and the gauge-free equations introduced in this work. }
\label{big1}
\end{center}
\end{figure*}

\section{Non-adiabaticity in BH inspirals} \label{evolution}

We now investigate the consequences of our new set of equations and their adiabatic limit, which is the same as Peters' equations. We perform this investigation numerically due to the coupled and non-linear nature of the expressions involved.

\subsection{Numerical evolution} \label{diabatic}

Figure~\ref{big1} shows the evolution of the orbital elements for an equal-mass binary initialized with $\bar{p}_0=20 M$, $\bar{e}_0=0.7$, $\bar{f}_0=-\pi$ and $\bar{\omega}_0=\pi$, employing the three methods described in Secs. ~\ref{lagrangian},~\ref{peters}, and ~\ref{ourmethod}.
We integrate from $\bar{t}_0=0$ to  $\bar{t}_{\rm f}=12933 M$, where the binary considered reaches a semi-latus rectum of $\bar{p}_{\rm f}=15 M$.
 The same initial values are used when comparing against Peters' equations, since at orders lower than 2.5PN the barred quantities coincide with their osculating counterparts.
For the osculating equations, we convert the initial conditions using the mappings of Eqs.~\eqref{pbars}, \eqref{ebars}, \eqref{wbars}, and \eqref{tbars} for the three sets of gauge parameters introduced in Sec. \ref{sec:rr}: the Harmonic gauge, the Burke-Thorne gauge, and the Sch\"{a}fer gauge.
This ensures that the initial conditions for the osculating equations are identical to those used for our new equations for semi-latus rectum, eccentricity, and true anomaly but show slight differences for $\omega_0$ and $t_0$, which we summarize in Table \ref{tab2}.%
\begin{table}[h!]
\centering
\renewcommand{\arraystretch}{1.5} 
\begin{tabular}{c|cc}
Gauge &  $ c^{5} \Delta \omega_0$ & $c^{5}  \Delta t_0$ \\
\hline
Harmonic   & $2.7\times 10^{-3}$ & $1.7\times 10^{-1}$ \\
Burke-Thorne  & $1.9\times 10^{-3}$ & $7.2\times 10^{-2}$ \\
Sch\"{a}fer & $2.3\times 10^{-3}$ & $1.2\times 10^{-1}$ \\
\end{tabular}
\caption{Difference between $\bar{\omega}_0$, $\bar{t}_0$ and $\omega_0$, $t_0$ for the BH system described in Sec.~\ref{diabatic}. The quantities $\omega_0$, $t_0$  are calculated from the initial values of the characteristic quantities as specified in Eq.~\eqref{wbars} and \eqref{tbars}.}
\label{tab2}
\end{table}

Our method, along with the Lagrangian planetary equations, successfully reproduces the %
{non-adiabatic} emission of GWs: the orbital elements, specifically $p$ and $e$, evolve in a step-like fashion, with drops occurring at pericenter passages where GW emission is the strongest. In contrast, Peters' equations capture only the secular evolution of these elements. %
At the same time, different gauge choices lead to distinct features in the evolution and, crucially, these differences lack %
a proper physical meaning. Among the four orbital elements, the evolution of $\omega$, which in our case is absent as described in Sec.~\ref{ourmethod}, %
is most prominently affected by these ambiguities. We emphasize once again that the evolution obtained using Eqs. \eqref{dpbdtb}–\eqref{dwbdtb} cannot be reproduced by any choice of  $\alpha$ and $\beta$.

\begin{figure*}[p]
\begin{center}
\includegraphics[width=\textwidth]{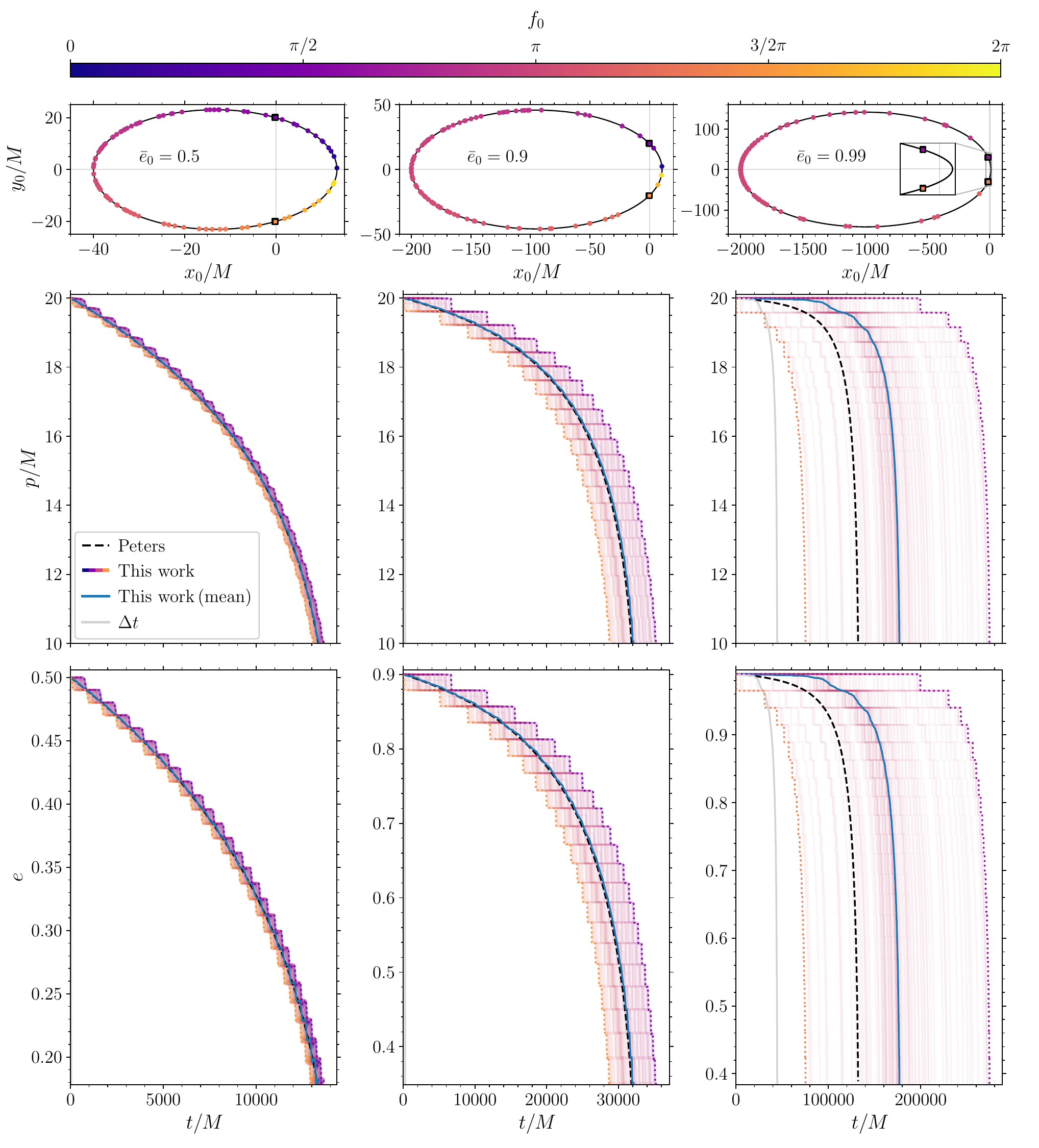} 
\caption{Evolution of the semi-latus rectum for binary systems with different initial eccentricities:  $\bar{e}_0=0.5$ (left panel),  $\bar{e}_0=0.9$ (central panel) and   $\bar{e}_0=0.99$ (right panel), all with the same initial semi-latus rectum $\bar{p}_0=20 M$ and concluding their evolution at $\bar{p}_0 = 10M$. For each pair ($\bar{p}_0,\bar{e}_0$), we sampled 100 points along the initial elliptic orbit (top panels) and used the corresponding true anomaly as the initial condition for the evolution. The larger panels illustrate the evolution obtained using the equations presented in Sec. \ref{ourmethod} (solid curves, colored according to the initial true anomaly) compared to Peters' equations (dashed black curves), which are insensitive to the initial starting point along the orbit. The mean evolution derived from our equations is shown in blue. The evolutions corresponding to the initial true anomalies marked by squares define the upper and lower boundaries enclosing all considered evolutions and are shown with dashed curves.}%
\label{big2}
\end{center}
\end{figure*}

\subsection{Breakdown of the adiabatic approximation} \label{diabatic2}

Figure~\ref{big1} shows that the orbital element evolution derived from Peters' equations aligns, on average, with the osculating evolution. Notably, no significant time shift is apparent, and all the examined evolutions reach approximately the same final values of eccentricity and semi-latus rectum at approximately the same time. This observation indicates that, while Peters' equations do not capture the finest details of the evolution, they provide a reliable treatment.

Conceptually, as highlighted in Sec.~\ref{peters}, orbital averaging should not be performed when the orbital parameters change on a timescale comparable to or shorter than the orbital period, i.e., when $\tau_{\rm rr} \lesssim \tau_{\rm orb}$.  In Fig.~\ref{big2}, we illustrate the consequences of this orbital-averaging breakdown on the evolution of $p$ and $e$.  As in Fig.~\ref{big1}, we initialize three binaries with the same semi-latus rectum $\bar{p}_0=20 M$ but different eccentricities $\bar{e}_0=(0.5,0.9,0.99)$. We evolve these systems until they reach a final separation of $\bar{p}_{f}=10M$, %
which we set as a conservative limit for our formalism and, more generally, for methods relying on the PN approximation \cite{2006PhRvD..74j4005B, 2009PhRvD..79h4010C,2009PhRvD..80h4043B}.
 Given a pair of initial conditions ($\bar{p}_0,\bar{e}_0$), we compute one hundred evolutions  with our %
 non-adiabatic equations, each with a different initial true anomaly $\bar{f}_0$, and taking into account the non-uniform probability of finding a body along an eccentric orbit (that is Kepler's area law). %

The mean of these evolutions is directly comparable to the evolution predicted by Peters' equations (which do not depend on the initial true anomaly). 
The time required for each system to reach the end of its evolution varies depending on the initial orbital positions.\footnote{We choose initial conditions such that $\bar{f}(t=0) = \bar{f}_{0}$, but on a Keplerian orbit, different values of $\bar{f}_{0}$ technically corresponds to different times by Kepler's equation. However, this formally reduces to an overall time shift in the solutions, which one always has the freedom to choose without affecting the underlying physics. If one corrects this, the spread in Fig.~\ref{big2} collapses to the average.} In other words, BH binaries evolve at different rates depending on the positions of the two bodies along the same initial orbit. %
Second, when the eccentricity is small to moderate ($e_0=0.5$ and $0.9$), the mean evolution obtained with our formalism is largely consistent with that of Peters'. This is not true in the highly eccentric case ($e_0=0.99$), where the evolution computed with Peters' equations is appreciably different compared to the mean of our non-adiabatic evolution.

\begin{figure*}[t]
\begin{center}
\includegraphics[width=0.95\textwidth]{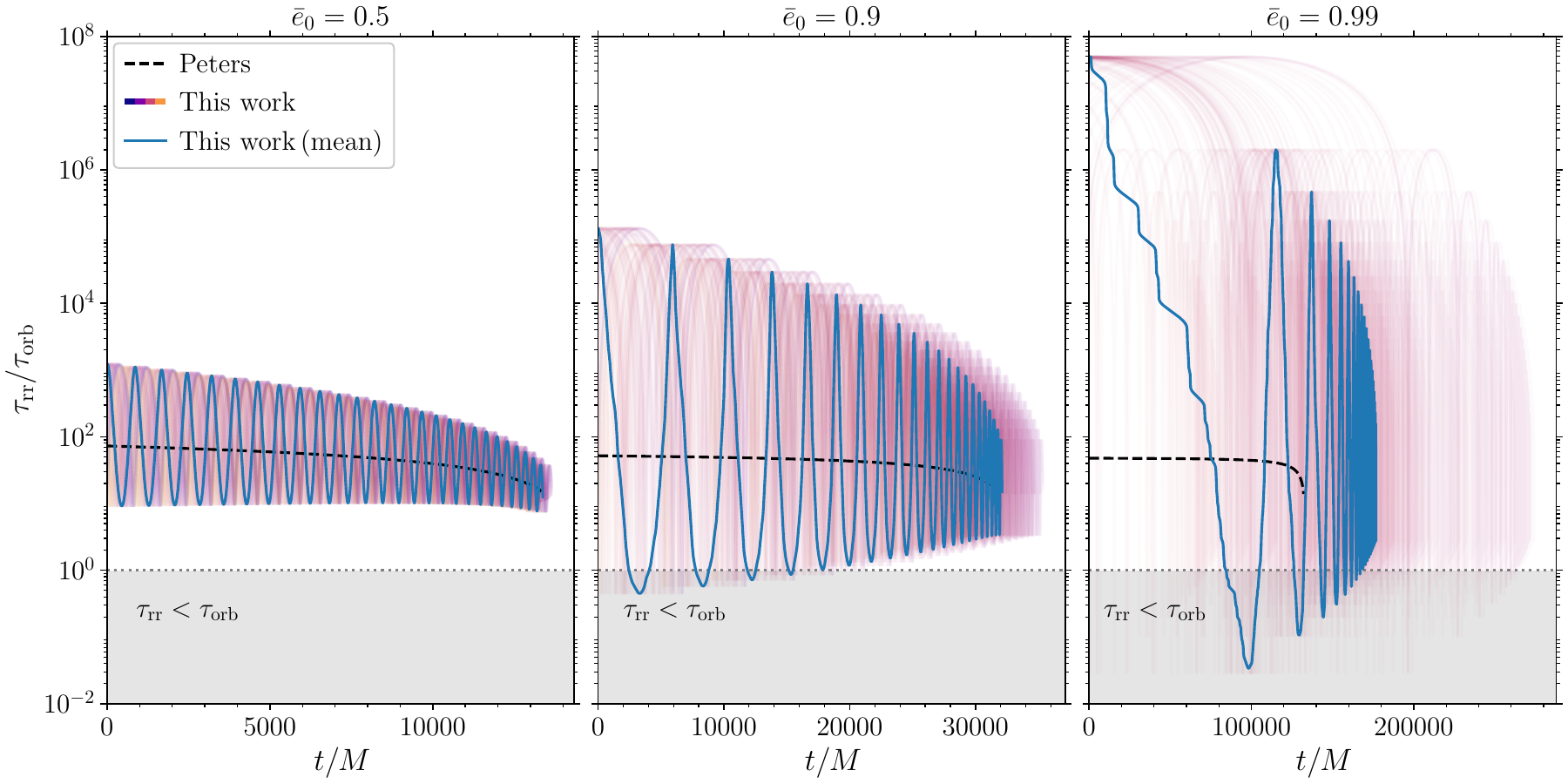} 
\caption{Evolution of the ratio $\tau_{\rm rr}/\tau_{\rm orb}$ for binary systems with different initial eccentricities: $\bar{e}_0 = 0.5$ (left panel), $\bar{e}_0 = 0.9$ (central panel), and $\bar{e}_0 = 0.99$ (right panel), all starting with the same initial semi-latus rectum $\bar{p}_0 = 20M$. The solid blue curves show the ratio's evolution, computed using the median values of $\bar{p}$, $\bar{e}$, and $\bar{f}$. The curves in a lighter shade show the timescale ratio for each of the evolutions presented in Fig.~\ref{big2}. The dashed black is computed using Peters' equations. The orbit-averaging procedure breaks down in the grey region where $\tau_{\rm rr} < \tau_{\rm orb}$.}
\label{timescaleexp}
\end{center}
\begin{center}
\includegraphics[width=0.95\textwidth]{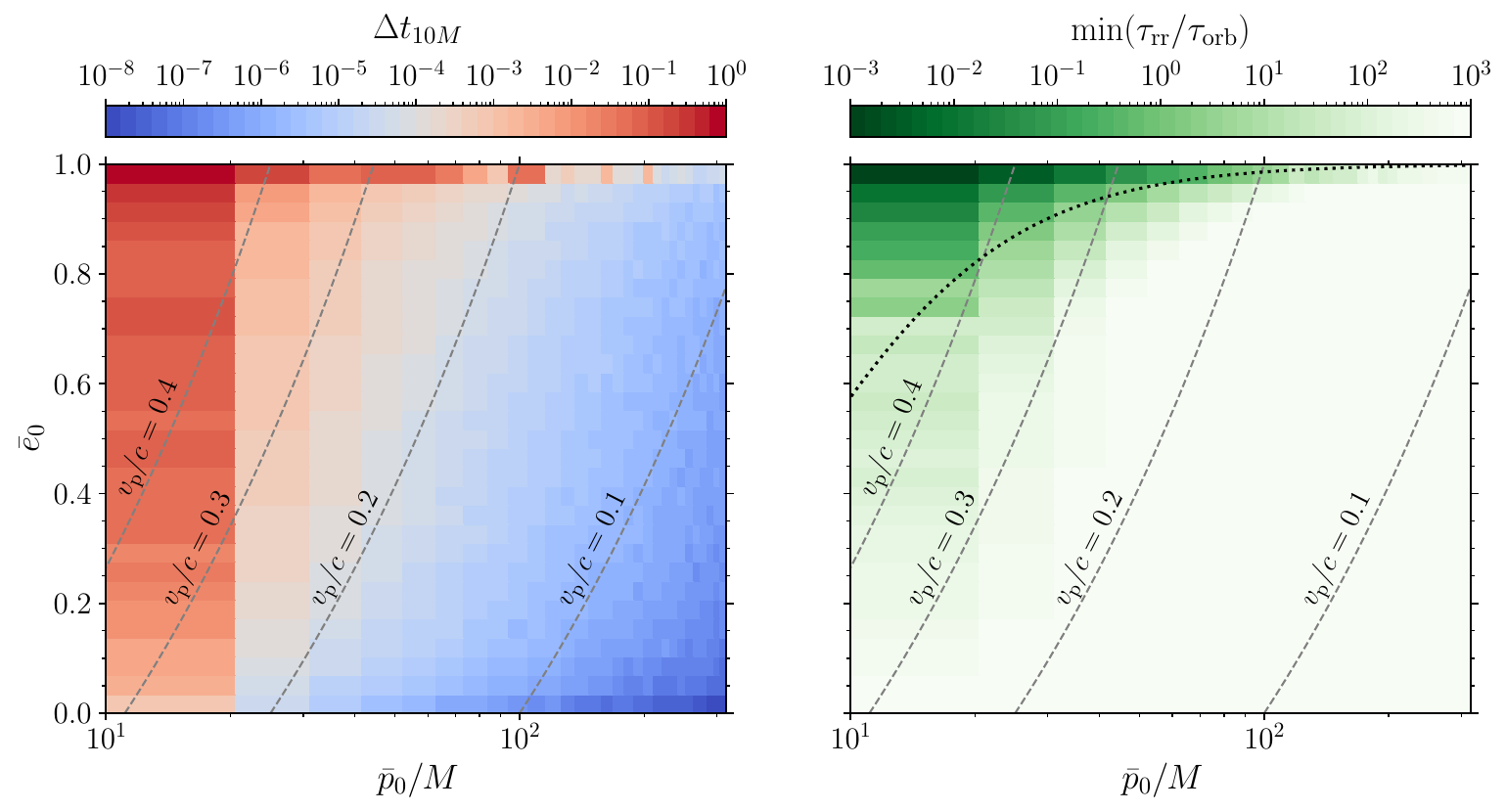} %
\caption{Breakdown of the orbital-average approximation in eccentric binary evolution across the parameter space. The left panel shows the normalized difference of the inspiral time $\Delta t_{\rm 10 M}$ between our formalism and Peters' as a function of the initial semi-latus rectum $\bar p_0$ and eccentricity $\bar e_0$. %
Specifically, we sample 100 initial values of the true anomaly for each pair ($p_0, e_0$). The right panel shows the timescale ratio $\tau_{\rm rr} / \tau_{\rm orb}$ for the mean evolution of these binaries. The dotted black curve indicates the separatrix of the timescale ratio, i.e., when its minimum along the evolution is equal to one. This is approximated with the condition $\bar{f}_0=\pi$. For reference, we report with dashed grey curves, contours of constant pericenter velocity $v_{\rm p}$ relative to the speed of light.} 
\label{dd}
\end{center}
\end{figure*}

To understand the reason for such divergence, Fig.~\ref{timescaleexp} illustrates the evolution of the ratio $\tau_{\rm rr }/\tau_{\rm orb}$ for the same integrations of  Fig.~\ref{big2}, measured using the single evolutions, their mean, as well as Peters' equations. It is important to note that, while $\tau_{\rm orb}$ is independent of the position along the orbit [see Eq.~\eqref{torb}], $\tau_{\rm rr}$ from Eq.~\eqref{trr} depends on the true anomaly, leading to oscillations that correspond to the variations in the orbital elements. In the case of small eccentricity, the timescale hierarchy is always well respected. As the eccentricity increases, the system crosses the $\tau_{\rm rr}=\tau_{\rm orb}$ boundary somewhere during its evolution. As expected, this occurs at the periastron passage, where most of the gravitational radiation is emitted.%

\subsection{Parameter-space exploration} \label{diabatic3}

In Fig.~\ref{dd}, we extend these calculations by considering a larger set of initial values with $\bar{p}_0 \in [10M, 10^{2.5}M] $ and $\bar{e}_0 \in [0, 1)$. For each $(\bar{p}_0, \bar{e}_0)$, we evolve the binary using Peters' equations and our radiation-reaction-gauge-free, %
non-adiabatic equations until it reaches $\bar{p}_{\rm f} = 10 M$. Since the last stable orbit corresponds to a semi-latus rectum of $p_{\rm LSO}= 2M (3+e)$, the evolutions reported remain within the regime where PN theory can still formally be defined, even in the most eccentric cases considered.  As in Fig~\ref{big2}, we select one hundred values of $\bar{f}_0$ for each $(\bar{p}_0, \bar{e}_0)$, and compute the mean over the resulting evolutions. 
We then extract the time taken by the binary to complete the evolution, denoting the value obtained using Peters' equations as $t_{\rm 10 M}^{\rm P}$ and the averaged value recovered with our non-adiabatic framework as $t_{\rm 10 M}^{\rm NA}$.
Specifically, we report the following quantity as a function of $(\bar{p}_0, \bar{e}_0)$:
\begin{align}\label{deltat10}
\Delta t_{\rm 10 M} = \frac{|t_{\rm 10 M}^{\rm P} - t_{\rm 10 M}^{\rm NA}|} {t_{\rm 10 M}^{\rm P}+t_{\rm 10 M}^{\rm NA}}   \in [0,1)\,.\end{align}

The left panel of Fig.~\ref{dd} shows that $\Delta t_{\rm 10 M} \sim 1$ when the eccentricity is high and the semilatus rectum is small. Conversely, in regions where the binary separation is large and the eccentricity is small, $\Delta t_{\rm 10 M}$ drops almost to zero, which implies Peters' equations can be used reliably.

A symptom of the breakdown of the orbit-averaged approximation, as already noted in Sec.~\ref{diabatic2}, is the inversion of the timescale hierarchy, i.e., $\tau_{\rm rr} \ll \tau_{\rm orb}$. Crucially,  for evolutions carried out with our formalism, the timescale $\tau_{\rm rr}$ depends on $f$, meaning that knowing the initial values $(p_0, e_0)$ is insufficient to determine whether the orbit-averaged approximation is valid. The right panel of Fig.~\ref{dd} shows the minimum value of the timescale ratio $\tau_{\rm rr}/\tau_{\rm orb}$ along the inspiral, calculated using the mean evolution. This minimum occurs at the first pericenter passage, revealing that regions where the timescale hierarchy is inverted, or equivalently where the ratio approaches unity, correspond to regions in the parameter space where the orbit-averaged approximation fails, resulting in $\Delta t_{\rm 10 M} \to 1$. The separatrix delineating the inversion of the timescale ratio is approximately given by the condition $\bar{f}_0=\pi$. %
Finally, let us note that for scenarios with moderate to high initial eccentricities, the timescale hierarchy is restored in the later stages of evolution, corresponding to a significant decrease in eccentricity.%

\begin{figure}[ht!]
\begin{center}
\includegraphics[width=\columnwidth]{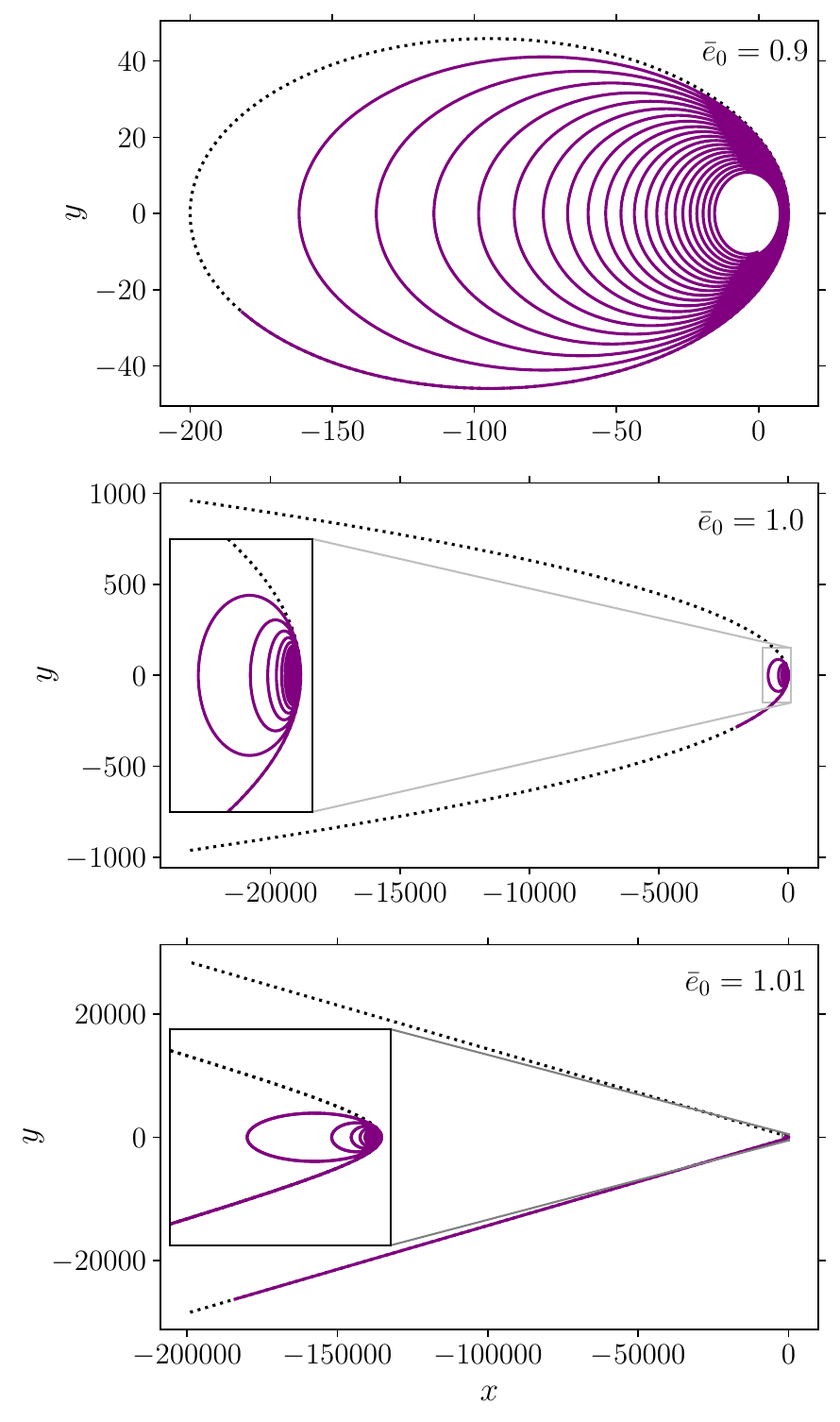} 
\caption{Time evolution of three binaries initially evolving on elliptical (top panel, $\bar{e}_0=0.9$), parabolic (middle panel, $\bar{e}_0=1$), and hyperbolic (bottom panel, $\bar{e}_0=1.01$) orbits. For all cases, we set $\bar{p}_0=20 M$, $\bar{f}_{0}=-3 $,  $\bar{\omega}_{0}=0$ and integrate up to $\bar{p}_{\rm f}=10 M$. The dotted black curves indicate the orbits described by the initial conditions. The purple curves are the non-adiabatic evolution under GW emission.}
\label{3ecc}
\end{center}
\end{figure}

The non-adiabatic formalism we propose, similarly to the osculating equations, can also handle parabolic $(e=1)$ and hyperbolic $(e>1)$ orbits. Figure \ref{3ecc} presents three examples generic binary systems, all starting with $\bar{p}_0=20 M$, and considering $\bar{e}_0=0.9$ (top panel), $\bar{e}_0=1$ (middle panel), and $\bar{e}_0=1.01$ (bottom panel). We chose an initial value true anomaly in the proximity of $ - \pi$ (where parabolic and hyperbolic orbits are singular), specifically setting at $\bar{f}=-3$ and $\bar{\omega}_{0}=0$. We evolve these binaries until they reach $\bar{p}_{\rm f}=10 M$. In both the parabolic and hyperbolic cases, binaries follow their respective unbound orbits before undergoing their first pericenter passage. During this event, a sufficient amount of energy is emitted via GW, causing the orbits to decay into bound elliptical trajectories.

\subsection{Gauge effects} \label{Gauge effects}

\begin{figure*}
\includegraphics[width=0.8\textwidth]{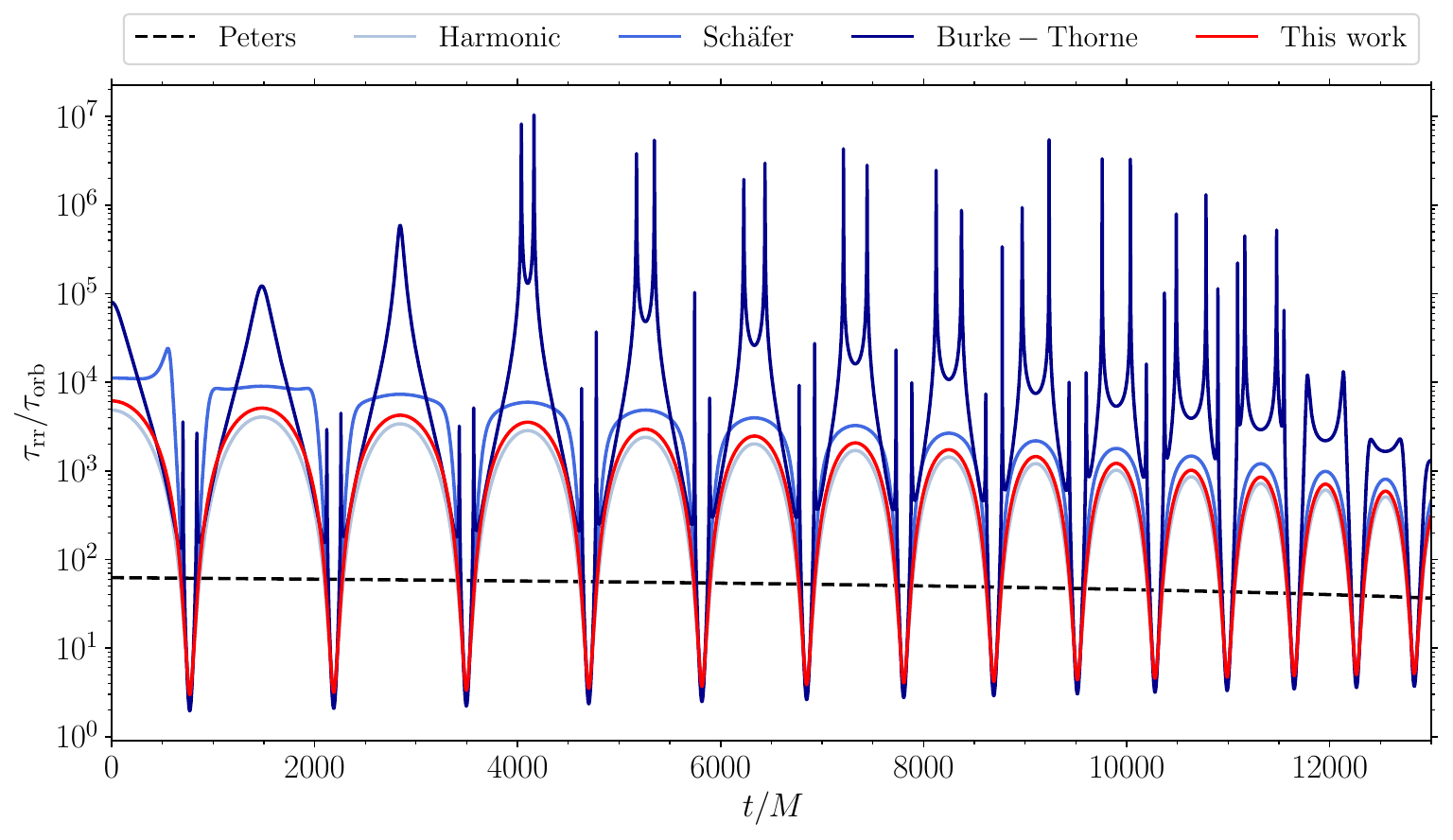} 
\caption{%
Evolution of the timescale ratio $\tau_{\rm rr}/\tau_{\rm orb}$ for the binary systems shown in Fig.~\ref{big1}. The solid curves in different shades of blue are computed using the Lagrangian planetary equation with three sets of gauge parameters $(\alpha, \beta)$: the harmonic gauge, the Schäfer gauge, and the Burke-Thorne gauge (for lighter to darker blue). The solid red curve is computed using our gauge-free equations. Finally, the dashed black curve is computed using Peters' orbit-averaged equations. }
\label{gaugeproof2}
\end{figure*}
An additional example of the effects induced by the freedom to choose gauge parameters in the Lagrange planetary equations is illustrated in Fig.~\ref{gaugeproof2}. We show the evolution of the timescale ratio $\tau_{\rm rr}/\tau_{\rm orb}$ for the binary system reported in Fig.~\ref{big1}, calculated using the Lagrangian planetary equations with three different gauge choices (see Sec. \ref{diabatic}), Peters' equations, and our non-adiabatic formalism.

When using the Lagrangian planetary equations, this timescale ratio depends significantly on the choice of gauge, with prominent spurious features that vary over time. These variations are, once more, gauge artifacts and do not describe the physics of BH binaries. Among the three gauge choices, the harmonic gauge yields the smoothest evolution, resulting in evolutions that are somewhat closer to those predicted by our formalism (see also Fig.~\ref{big1}).

This sensitivity to gauge choices makes the Lagrangian planetary equations impractical for evaluating the breakdown of the orbital averaging approximation, as discussed in Sec. \ref{diabatic}. In contrast, our formalism demonstrates a consistent behavior, free from spurious effects, while still capturing non-adiabatic effects.

\section{Discussion}\label{conclusion}

\subsection{Key findings}

Eccentric binaries are among the most anticipated GW sources, with the potential to provide critical insights into the formation channels of BH binaries \cite{2021ApJ...921L..43Z,  Zeeshan:2024ovp}. These systems are key targets for current ground-based detectors and, in particular, for future space-based observatories like LISA \cite{2018PhRvD..97j3014S,2018MNRAS.481.5445S}. However, their accurate modeling remains challenging due to the non-adiabatic nature of GW emission, which significantly influences their orbital dynamics and long-term evolution.

In this work, we re-investigated the foundation of BH-binary dynamics in the PN regime and presented a non-adiabatic framework that is not affected by radiation-reaction gauge parameters, addressing the limitations of established methods such as the adiabatic approximation~\cite{1964PhRv..136.1224P} and the Lagrangian planetary formalism~\cite{LincolnWill}. The key result of this paper is presented in Eqs. \eqref{dpbdtb}–\eqref{dfbdtb}.

By eliminating the dependence on radiation-reaction gauge choices and avoiding orbit averaging, our formalism ensures a strictly physical, and therefore possibly observable, interpretation of the binary parameters and their evolution. This approach remains valid throughout the entire parameter space $(\bar{p}, \bar{e})$ where the PN approximation is applicable, including the parabolic and hyperbolic limits.

With these new equations, we quantify the regime of applicability of the popular Peters' equations, specifically that systems must satisfy the timescale condition $\tau_{\rm rr}\gg \tau_{\rm orb} $. Even if such a condition is widely known, our investigation highlights some non-trivial points. In particular, we find that:
\begin{itemize}
\item[(i)] An inappropriate usage of Peters' equations can result in significant miscalculation of the inspiral time of BH binaries, most crucially for binaries with large initial eccentricities and small semi-latus recta.
\item[(ii)]  Initial conditions on eccentricity and semi-latus rectum (or equivalently semi-major axis) alone are insufficient to predict whether the orbit-averaging approximation breaks down. Instead, one must evaluate the timescale separation at the first periastron passage, which depends on the initial value of the true anomaly.
\end{itemize}

\subsection{Implications}

A  detailed discussion of the astrophysical implications of our new set of equations is left for future work, including potential systematics in waveform modeling and astrophysical inference for eccentric BH binaries. %
For now, we provide some order-of-magnitude estimates of the differences in predictions obtained using our formalism and the common orbit-averaged approach.

Considering the three scenarios shown in Fig.~\ref{big2}, we assign physical units by setting a total mass of $60 M_{\odot}$ and an initial semi-latus rectum of $\bar{p}_0=2 \times 10^{-3} R_{\odot}$. Table~\ref{finalnumbers} the times required (and their differences) to complete the evolutions according to both Peters' equations and our new formalism. We consider the two extreme cases, i.e., the fastest and slowest evolution, corresponding to $\bar{f}_0 \approx \pi/2$ and $\bar{f}_0 \approx 3\pi/2$, respectively, together with the mean evolution.
For the most eccentric case, the evolution time calculated when $\bar{f}_0 \approx 3\pi/2$ is larger than the value predicted by Peters' equations by about a factor of 2. %
  
\begin{table}[t]
\centering
\renewcommand{\arraystretch}{1.5} 
\begin{tabular}{c|ccccc}
$\bar{e}_0$ & $t_{\rm 10M}^{\rm P}$ & $t_{\rm 10M}^{\rm NA, {\pi/2}}$ & $t_{\rm 10M}^{\rm NA, {3\pi/2}}$ & $t_{\rm 10M}^{\rm NA}$ & $\Delta t_{10M}$ \\ 
\hline
$0.5$  & $3.97$  & $4.05\,(0.09)$  & $3.88\,(0.08)$  & $3.97\,(0.01)$  & $0.001$  \\
\hline
$0.9$  & $9.44$  & $10.47\,(1.03)$ & $8.55\,(0.89)$  & $9.52\,(0.08)$  & $0.004$  \\
\hline
$0.99$ & $39.18$ & $81.34\,(42.16)$ & $22.30\,(16.87)$ & $52.41\,(13.23)$ & $0.144$  \\
\end{tabular}
\caption{Times, expressed in seconds, required to complete the evolutions shown in Fig.~\ref{big2} for a non-spinning, equal-mass BH binary with $M=60 M_{\odot}$ and $\bar{p}_0=2 \times 10^{-3} R_{\odot}$, calculated using Peters' equations ($t_{\rm 10M}^{\rm P}$) and our formalism ($t_{\rm 10M}^{\rm NA}$) for $\bar{f}_0 \approx \pi/2$, $\bar{f}_0 \approx 3\pi/2$, and the mean evolution. In parentheses, we report the absolute difference between the inspiral times of our evolution and those from Peters'. Finally, the quantity $\Delta t_{10M}$ is defined in Eq.~(\ref{deltat10}).}
\label{finalnumbers}
\end{table}

\subsection{Future prospects}

This analysis could not be reliably performed using the Lagrangian planetary equations due to the presence of spurious and uncontrolled features in the evolution of the orbital parameters. However, it is worth stressing that the evolutions obtained when assuming the harmonic gauge produce the most consistent results are similar to those from our gauge-free method. 
To assess whether the proposed representation of the dynamics leads to a formally gauge-invariant description (in the sense of independence from gauge parameters, as is the case for globally defined quantities such as the total energy), one must now compute the energy and angular momentum fluxes. Due to the complexity of the calculation, we defer this analysis to future work.

This work is grounded in and directly compared to the osculating formalism, which incorporates the Newtonian and 2.5PN terms only in both the equations of motion and the definitions of energy and angular momentum. 
While terms up to 4.5 PN are available, including both conservative and dissipative terms \cite{2002PhRvD..65j4008P, 2006PhRvD..73l4012K}, deriving Lagrangian planetary equations consistent with these higher orders requires a more detailed and careful treatment than that outlined in Sec.~\ref{lagrangian} as suggested in Ref. \cite{2004PhRvD..70f4028D}.
Most importantly, incorporating corrections to the dissipative terms at 3.5PN order introduces six new gauge parameters \cite{1995PhRvD..52.6882I, 2025CQGra..42f5015B}, while employing 4.5PN corrections adds twelve additional gauge parameters.
Although our method can, in principle, be extended to higher PN orders, including these corrections is non-trivial and would require a significantly more convoluted derivation.  We leave such extensions to future work. 
 Other non-adiabatic treatments of the dynamics of eccentric BH binaries have been presented in e.g. Ref.~\cite{2021PhRvD.103f4066T, 2023PhRvD.107j3040T}, and a detailed comparison with those methods is another interesting avenue for future work. 
Finally, this study considers non-spinning black hole binaries. Some of the authors \cite{2023PhRvD.108l4055F, 2024PhRvD.110f3012F} have shown a correlation between the evolution of spin direction and eccentricity. Investigating how this new formalism affects this interplay, and vice versa, as well as the influence of spin on the evolution of eccentricity, is another important avenue that requires further careful study.

 \acknowledgements

We thank Chris Moore, Clifford Will, and Isobel Romero-Shaw for discussions. 
G.F., N.L., D.G., and M.B. are supported by 
ERC Starting Grant No.~945155--GWmining, 
Cariplo Foundation Grant No.~2021-0555, 
MUR PRIN Grant No.~2022-Z9X4XS, 
MUR Grant ``Progetto Dipartimenti di Eccellenza 2023-2027'' (BiCoQ),
and the ICSC National Research Centre funded by NextGenerationEU. 
D.G. is supported by MSCA Fellowships No.~101064542--StochRewind
and No.~101149270--ProtoBH.
Computational work was performed at CINECA with allocations 
through INFN and Bicocca.

\bibliography{gauge_free}

\end{document}